\documentclass{emulateapj}
\usepackage{natbib,aas_macros}

\newcommand{\vk}{v_{\rm K}}
\newcommand{\va}{v_{\rm A}}
\newcommand{\Pth}{P_{\rm th}}
\newcommand{\mbh}{M_{\rm BH}}
\newcommand{\msun}{M_{\odot}}
\newcommand{\mbhn}{M_{\rm BH,7}}
\newcommand{\kb}{k_{\rm B}}
\newcommand{\delt}{\frac{\partial}{\partial t}}
\newcommand{\delp}{\frac{\partial}{\partial p}}
\newcommand{\rs}{R_{\rm S}}

\newcommand{\tcool}{t_{\rm cool}}

\newcommand{\taccel}{t_{\rm accel}}
\newcommand{\tfall}{t_{\rm fall}}
\newcommand{\tdiff}{t_{\rm diff}}
\newcommand{\tsync}{t_{\rm sync}}
\newcommand{\finj}{\dot F_{\rm inj}}
\newcommand{\sigmat}{\sigma_{\rm T}}
\newcommand{\sigmapeak}{\sigma_{\rm pk}}
\newcommand{\epeak}{\bar{\epsilon}_{\rm pk}}
\newcommand{\Kpeak}{K_{\rm pk}}

\newcommand{\etacr}{\eta_{\rm cr}}
\newcommand{\mdedd}{\dot M_{\rm Edd}}
\newcommand{\ledd}{L_{\rm Edd}}
\newcommand{\ljet}{L_{\rm jet}}
\newcommand{\epeq}{E_{p,\rm eq}}
\newcommand{\ecut}{E_{\gamma,\rm cut}}
 \newcommand{\gammapeq}{\gamma_{p,\rm eq}}
\newcommand{\epmax}{E_{p,\rm pk}}
\newcommand{\enumax}{E_{\nu,\rm pk}}

\newcommand{\pinj}{p_{\rm inj}}
\newcommand{\lmax}{L_{\rm max}}
\newcommand{\lmin}{L_{\rm min}}
\newcommand{\mdmax}{\dot m_{\rm max}}
\newcommand{\mdmin}{\dot m_{\rm min}}
\newcommand{\mdjet}{\dot M_{\rm jet}}

\newcommand{\kmin}{k_{\rm min}}
\newcommand{\zmax}{z_{\rm max}}

\shorttitle{Neutrino and CR Emission from RIAFs in LLAGN}
\shortauthors{Kimura, Murase, \& Toma}

\begin{document}

\title{Neutrino and Cosmic-Ray Emission and Cumulative Background from\\
Radiatively Inefficient Accretion Flows in Low-Luminosity Active Galactic Nuclei}
\author{Shigeo S. Kimura\altaffilmark{1,2}, Kohta Murase\altaffilmark{3,4}, and Kenji Toma\altaffilmark{1,2}}
\affil{$^1$Astronomical Institute, Tohoku University, Sendai 980-8578, Japan}
\affil{$^2$Frontier Research Institute for Interdisciplinary Sciences, Tohoku University, Sendai 980-8578, Japan}
\email{kimura@vega.ess.sci.osaka-u.ac.jp}
\affil{$^3$Hubble Fellow---Institute for Advanced Study, Princeton, New Jersey 08540, USA}
\affil{$^4$Center for Particle and Gravitational Astrophysics; Department of Physics; Department of Astronomy \& Astrophysics; The Pennsylvania State University, University Park, Pennsylvania, 16802, USA}
\email{shigeo@astr.tohoku.ac.jp}

\begin{abstract}
We study high-energy neutrino and cosmic-ray (CR) emission from the cores of low-luminosity active galactic nuclei (LLAGN).  In LLAGN, the thermalization of particles is expected to be incomplete in radiatively inefficient accretion flows (RIAFs), allowing the existence of non-thermal particles. 
In this work, assuming stochastic particle acceleration due to turbulence in RIAFs, we solve the Fokker-Planck equation and calculate spectra of escaping neutrinos and CRs.
 The RIAF in LLAGN can emit CR protons with $\gtrsim10$~PeV energies and TeV--PeV neutrinos generated via $pp$ and/or $p\gamma$ reactions.
We find that, if $\sim1$\% of the accretion luminosity is carried away by non-thermal ions, the diffuse neutrino intensity from the cores of LLAGN may be as high as $E_\nu^2\Phi_\nu\sim3\times{10}^{-8}~{\rm GeV}~{\rm cm}^{-2}~{\rm s}^{-1}{\rm sr}^{-1}$, which can be compatible with the observed IceCube data.  This result does not contradict either of the diffuse gamma-ray background observed by {\it Fermi} or observed diffuse cosmic-ray flux.
 Our model suggests that, although very-high-energy gamma rays may not escape, radio-quiet AGN with RIAFs can emit GeV gamma-rays, which could be used for testing the model.
 We also calculate the neutron luminosity from RIAFs of LLAGN, and discuss a strong constraint on the model of jet mass loading mediated by neutrons from the diffuse neutrino observation. 
\end{abstract}


\section{Introduction}
The IceCube collaboration reported a discovery of extraterrestrial neutrinos with deposited energies ranging from 30 TeV to a few PeV, and the significance now exceeds $5\sigma$ \citep{ice13sci,ice13prl,ice14}. The signals are likely to be astrophysical, and the consistency with isotropic distribution suggests extragalactic components, which is also supported by diffuse gamma-ray data \citep{mal13,am14}.  Significant clustering has not been observed, and the origin of IceCube neutrinos is a new big mystery even though some early models can match the observed data within large model uncertainties \citep{anc+04,ste05,lw06,mur+06,gz07,min08,kot+09}. One of the popular possibilities is neutrino emission from cosmic-ray (CR) reservoirs.
Star-forming galaxies, including starbusrt galaxies, may explain the IceCube data, and various possibilities have been speculated to have $\gtrsim10-100$~PeV CR protons \citep{mal13,kat+13,tam14}.
Galaxy groups and clusters may also account for the data without violating gamma-ray limits.  While $\gtrsim10-100$~PeV CR protons can be supplied by active galactic nuclei (AGN), galaxies and galaxy mergers
\footnote{CR sources like AGN and galaxies strongly evolve as redshifts.}
as well as intergalactic shocks, hard spectral indices $s_\nu\sim2$ and contributions from low-mass clusters and groups would be needed~
\footnote{While neutrino and gamma-ray flux calculations by the recent work by \cite{zan+14} is actually consistent with \cite{min08,mia09} for the massive cluster shock case, the setup noted in \cite{mal13} has not been tested. Connecting individual massive cluster emission to diffuse backgrounds depends on models and underlying assumptions, and astrophysical uncertainty is still too large to cover all the relevant parameter space.}
\citep{mal13,km14}. However, due to multi-messenger constraints \citep{mal13}, the above scenarios may be challenged by the latest data implying steep indices $s_\nu\sim2.3-2.5$ \citep{ice14,ice15} and models that attribute the diffuse gamma-ray background to unresolved blazars. 

High-energy neutrino production in AGN has been of interest for many years \citep[e.g.,][]{pk83,ke86,ste+91,sp94,man95,ad01,am04,anc+04}. The most popular possibility is photohadronic ($p\gamma$) neutrino production in relativistic jets that are established as gamma-ray sites \citep[e.g.,][]{man95,ad01,mp01}. The diffuse neutrino intensity of radio-loud AGN is typically dominated by luminous blazars, in which external radiation fields due to accretion disk, broadline and dust emission are relevant \citep{mid14}. However, it has been shown that the simple inner jet model has difficulty in explaining the IceCube data, and additional assumptions are required \citep{dmi14,tgg14}.  Alternatively, high-energy neutrino emission could mainly come from the cores of AGN, and both of $pp$ \citep{bec+14} and $p\gamma$ \citep{ste13,win13,kst14} scenarios have been considered. In the latter case, it has been assumed that target photon fields come from the standard, Shakura-Sunyaev disk \citep{ss73}, but the disk temperature to account for $\lesssim300$~TeV neutrinos has to be higher than typical values \citep{dmi14}.   

One big issue in the AGN core models is how to have non-thermal protons in such inner regions. Electric field acceleration has been discussed \citep{lev00}, but the formation of a gap above the black hole is not clear in the presence of copious plasma. The shock dissipation may also occur in accretion flows \citep{brs90,am04,bdl08}, although efficient shock acceleration is possible when the shock is not mediated by radiation. When the accretion rate is high enough to form the standard disk \citep{ss73} or the slim disk \citep{abr+88}, protons and electrons should be thermalized via the Coulomb scattering within the infall time. This indicates that turbulent acceleration is unlikely in the bulk of the accretion flow, although possible non-thermal proton acceleration in the corona of standard disks has also been discussed \citep[e.g.,][]{dml96,rvv10}. 
 
In this work, we consider the possibility of high-energy neutrino emission from low-luminosity AGN (LLAGN), which has not been discussed in light of the IceCube data. It has been considered that LLAGN do not have standard or slim disks, since their spectra show no blue bump \citep{ho08}. Instead, LLAGN are believed to have the radiation inefficient accretion flows \citep[RIAFs,][]{ny94}, in which plasma can be collisionless with low accretion rates, allowing the existence of non-thermal particles \citep{mq97,tt12}. In the unified picture of AGN, BL Lac objects correspond to LLAGN viewed from an on-axis observer, whereas quasar-hosted blazars correspond to highly accreted AGN with optically thick disks.
 
Turbulent magnetic fields play an important role for the angular momentum transport in accretion disks, and the magnetic rotational instability (MRI) has been believed to be responsible for the effective viscosity \citep[e.g.,][]{bh91,san+04}. Recent particle-in-cell simulations have shown that the magnetic reconnection occurring in the MRI turbulence generates non-thermal particles \citep{riq+12,hos13,hos15}, although these simulations follow the plasma scale structure that is much smaller than the realistic scale of the accretion flow. It would be also natural to expect stochastic acceleration in the presence of strong turbulence in RIAFs \citep[e.g.,][]{lyn+14}. 

Protons accelerated in RIAFs lead to gamma-ray emission via $pp$ and $p\gamma$ processes \citep{mnk97,nxs13}. In this work, we focus on neutrino and CR emission, motivated by the latest IceCube discovery. The acceleration efficiency is uncertain at present. \citet{ktt14} shows that high-energy particles do not affect the dynamical structure, except for the cases that they extract most of the released energy from RIAFs. This implies that the energy loss rate by high-energy protons is limited to several percents of $\dot M c^2$, where $\dot M$ is the mass accretion rate, but we show that it is still possible for LLAGN to make significant contributions to the cumulative neutrino background without violating existing observational limits.   

In this paper, we consider stochastic acceleration in RIAFs of LLAGN and suggest that they are potential sources of high-energy neutrinos. In Section \ref{sec:bg}, we set up physical states of RIAFs. In Section \ref{sec:spectra}, we formulate and calculate energy spectrum of the non-thermal protons inside a typical RIAF. The spectra of escaping protons and neutrinos from the RIAF are also presented in Section \ref{sec:spectra}. Then, the diffuse neutrino intensity is estimated and compared to the IceCube data in Section \ref{sec:diffuse}. We discuss several related issues such as the detectability in gamma rays in Section \ref{sec:discussion}, and summarize our results in Section \ref{sec:summary}.

\section{Physical Setup}\label{sec:bg}

We model emission from RIAFs with the one-zone approximation, where it is assumed that particles are accelerated only within some radius $R$. When one considers the structure of accretion disks, the multi-dimensionality is important in general. Nevertheless, we consider that our approach is enough as the first step to consider high-energy neutrino emission from RIAFs.

\subsection{Physical quantities of RIAFs}

RIAFs are the hot and rapid infall accretion flows. We set the radial velocity $v_r$, thermal proton density $n_p$, thermal pressure $\Pth$, and strength of magnetic fields $B$ of our RIAF model as follows,
\begin{eqnarray}
 v_r &=& \alpha \vk,\\
 n_p &=& \frac{\dot M}{2\pi R^2 v_r m_p},\\
 \Pth &=& n_p \frac{G\mbh}{3R} m_p,\\
 B &=& \sqrt{\frac{8\pi \Pth}{\beta}},
\end{eqnarray}
where $\alpha$ is the alpha parameter \citep{ss73}, $\vk=\sqrt{G\mbh/R}$ is the Keplerian velocity, $\dot M$ is the mass accretion rate, $\mbh$ is the mass of the super massive black hole (SMBH), and $\beta$ is the plasma beta parameter. We assume the scale height of the flow $H\sim R$. We normalize the radius and mass accretion rate as $r = R/\rs$ and $\dot m = \dot M/\mdedd$, respectively, where we use the Schwarzschild radius $\rs = 2G\mbh/c^2$ and the Eddington accretion rate $\mdedd = \ledd/c^2$. This makes
\begin{eqnarray}
 R &=& 2.95\times 10^{13}~r_{1} \mbhn ~{\rm cm}~,\\
 v_r &=& 6.7\times 10^{8}~r_{1}^{-1/2} \alpha_{-1}~{\rm cm~s^{-1}}~,\\
 n_p &=& 1.1\times 10^{9}~r_{1}^{-3/2} \alpha_{-1}^{-1} \mbhn^{-1} \dot m_{-2} ~{\rm cm^{-3}}~,\\
 B &=& 4.9 \times 10^{2}~r_{1}^{-5/4} \alpha_{-1}^{-1/2} \beta_{3}^{-1/2} \mbhn^{-1/2} \dot m_{-2}^{1/2}~{\rm Gauss}~, 
\end{eqnarray}
where $A_{n}=A/10^n$, except $M_{{\rm BH},n} = \mbh/(10^n M_{\odot})$ and $\beta_3 = \beta/3$. 
For reference, the accretion luminosity and Thomson optical depth are computed as 
\begin{eqnarray}
\dot M c^2 &=& 1.3\times 10^{43} \mbhn \dot m_{-2}   \rm ~ erg~s^{-1},\\
\tau_{\rm T} &=& n_p \sigmat R = 2.2\times 10^{-2}  r_1^{-1/2} \alpha_{-1}^{-1}\dot m_{-2},
\end{eqnarray}
where $\sigmat$ is the Thomson cross section.
If we consider small $r \lesssim 5$, $v_r$ and $n_p$ are quite different from above expression because the flow becomes supersonic and particles go into the SMBH quickly \citep{nkh97,ktt14}.
In this paper, we fix the parameters $\alpha=0.1$ and $r=10$ for demonstration.

\subsection{Thermal electrons and target photon fields}\label{sec:targetphoton}

The photomeson production is an important process of the neutrino and/or gamma-ray production. 
The sub-PeV and PeV neutrinos are generated by the interaction
between protons and photons of the orders of $E_p\sim 10^{16}-10^{17}$ eV and $E_\gamma\sim 1-10$ eV, respectively.
Thus, we estimate the target photon spectrum in RIAFs. The estimate of 
luminosity of LLAGN, $L_{E_\gamma}$, will be also used to calculate the diffuse 
neutrino flux in Section \ref{sec:diffuse}.

It has been suggested that in LLAGN, emission comes from a jet, an outer thin disk, and a RIAF \citep{nem+06,nse14}. In this paper, we consider only radiation from the RIAF because radiation from the jet and thin disk would be sub-dominant. We use the one-zone approximation and calculate the photon spectrum within acceleration radius $R$. Thermal electrons in the RIAF emit radiation through the synchrotron, bremsstrahlung, and inverse Compton scattering. 
We use fitting formulae of the emissivity of bremsstrahlung and synchrotron \citep{ny95}. Assuming the local thermodynamic equilibrium with Eddington approximation \citep{rl79}, we can get the photon fields from synchrotron and bremsstrahlung. This treatment consistently includes the synchrotron self absorption \citep{mmk97}. Using this photon fields as the seed photons, spectra of inverse Compton scattering are calculated.  See the Appendix for details of the calculation of target photon fields. 

To obtain the target photon spectra, we need to know the electron temperature. Since the relaxation time between electrons and protons in RIAFs is longer than the infall time $\tfall$ (see Section \ref{sec:plasma}), electrons would have different temperature from that of the protons \citep{tk85}.
The electron temperature is usually determined by an energy balance, $Q_+ = Q_-$,
where $Q_+$ and $Q_-$ is the heating rate and cooling rate of electrons, respectively. 
The mechanism of electron heating in RIAFs is determined by details of dissipation in collisionless plasma \citep{qg99,sha+07,how10}, but the accurate prescription for the turbulent heating is not well understood.
Here, we assume a simple heating prescription
where the heating rate of electrons are proportional to the accretion luminosity, i.e.,
\begin{equation}
  Q_+ = \delta_e\dot M c^2,
\end{equation}
where $\delta_e$ is the heating parameter that represents the fraction of energy that directly heats up electrons.
We consider the range of $10^{-3} \le \delta_e \le 5\times10^{-2}$ following to \citet{nse14}, and use $\delta_e\sim 3\times10^{-3}$ as a fiducial value
\footnote{This definition of $\delta_e$ is different from that in the previous works including \citet{nse14}. They define the heating parameter $\delta$ as $Q_+ = \delta Q_{\rm vis}$, where $Q_{\rm vis}$ is the viscous dissipation rate. Since the viscous dissipation rate is typically the order of $Q_{\rm vis} \sim 0.1 \dot M c^2$, the value of $\delta_e=0.05$ in our model corresponds to $\delta=0.5$ for the previous works.}.
The cooling rate is estimated from the total luminosity of target photons $Q_-=\int L_{E_\gamma}dE_\gamma$, where we use the assumption of optically thin limit. 
We calculate the equilibrium temperature, $\theta_{e,\rm eq}$, from the energy balance $Q_+=Q_- (\theta_{e,\rm eq})$ 
using the bisection method \citep{pre+92}.
Since the source of thermal energy is the released gravitational energy,
the electron temperature cannot exceed the virial temperature of electron-proton gas, $\theta_{e,\rm vir} = GM m_p/(9R)$.
Thus, the electron temperature is written as $\theta_e = {\rm min} (\theta_{e,\rm eq}, \theta_{e,\rm vir})$.
Then, we assume the distribution function of thermal electrons is the relativistic Maxwellian,
\begin{equation}
 N_e(\gamma_e) = n_e \frac{\gamma_e^2\beta_e \exp(-\gamma_e/\theta_e)}{\theta_e K_2(1/\theta_e)}, 
\end{equation}
where $n_e$ is the electron number density, $\beta_e$ and $\gamma_e$ are the velocity and the Lorentz factor of the thermal electrons, respectively, and $K_2(x)$ is the second modified Bessel function. We ignore effects of the pair production on the thermal component for simplicity, which gives $n_p = n_e$.
We do not consider non-thermal electrons because electrons have much shorter relaxation time than protons.
When $\dot m \gtrsim 10^{-4}$,
they become thermalized within infall time through the synchrotron self absorption process \citep{mq97}. 
For $\dot m \lesssim 10^{-4}$, the electrons seem to be non-thermal. 
However, we ignore the effect of non-thermal electrons 
because such low-luminosity objects are less important for the diffuse neutrino flux (see Section \ref{sec:diffuse}). 

We tabulate the values of $\theta_e$ for some models in Table \ref{tab:single}
(where the other resultant quantities will be introduced later). 
We fix the parameters $\alpha=0.1$, $\beta=3$, $r=10$, and $\delta_e=3\times10^{-3}$.
For the reference model A1, $\theta_e\sim2.0$ and it does not depend on $\mbh$ very much.
The high $\delta_e$ leads to high $\theta_e$ due to the high heating rate.
For the flows with lower density (lower $\dot m$) or weaker magnetic fields (higher $\beta$),
the electron temperature is higher because the cooling rate for a given $\theta_e$ is lower in such flows.

\begin{figure}
   \includegraphics[width=\linewidth]{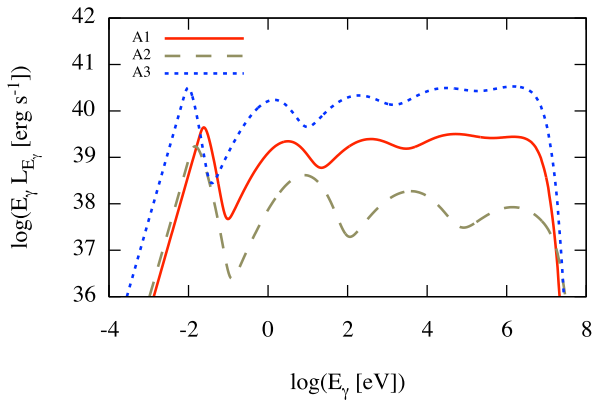}
   \centering
   \caption{Target photon spectra emitted by thermal electrons in RIAFs.
   The red-solid, the green-dashed, and blue-dotted lines show
   models A1 (reference), A2 (low $\dot m$), A3 (high $\mbh$), respectively.
   The target photon spectrum for model A4 is the same with that for A1. 
   \label{fig:bg-photon}}
   \end{figure}

Figure \ref{fig:bg-photon} shows target photon spectra in RIAFs for models A1, A2, and A3
(for models A4 and A5, see Section \ref{sec:spectra} and \ref{sec:diffuse}, respectively). 
The values of $\tau_{\rm T}$ and $L_X$ are tabulated in Table \ref{tab:single},
where $L_X$ is the X-ray luminosity in the 2-10 keV band. 
For model A1, the synchrotron component has a peak at $E_\gamma\sim 0.03$ eV. The thermal electrons scatter seed synchrotron photons efficiently, and make a few peaks from the infrared to soft X-ray range.  Multiple-scattered photons may make an almost flat spectrum for the hard X-ray range. The spectrum has a cutoff corresponding to the electron temperature. The inverse Compton scattering dominates over the bremsstrahlung in all the frequency range for A1.

The efficiency of inverse Compton scattering depends on the $y$ parameter, $y \sim \tau_{\rm T} \theta_e^2 \propto \dot m r^{-1/2} \alpha^{-1} \theta_e^2$, where $\tau_{\rm T}=n_e \sigmat R$ is the optical depth for Thomson scattering.
Although $\theta_e$ is higher as $\dot m$ is lower, it makes the $y$ parameter lower.
Thus, the spectrum by inverse Compton scattering is softer for lower $\dot m$. 
For A2, $y$ parameter is less than unity, but it still dominates over the bremsstrahlung in all the range.
The $y$ parameter is independent of $\mbh$ in our formulation. The high $\mbh$ makes luminosity higher due to high values of $R$ and $\dot M$. It also makes the synchrotron peak frequency low because of weak $B$.
The profile of the spectrum for A3 is similar to that for A1 but the luminosity for A3 is about ten times higher than that for A1.
When the electron temperature is higher with fixed $\dot m$, the $y$ parameter become higher.
Thus, the spectrum is harder for higher $\beta$ and higher $\delta_e$. 

\begin{table*}[tb]
\begin{center}
\caption{Model parameters and resultant physical quantities for the spectrum from a LLAGN  \label{tab:single}}
\begin{tabular}{|c|ccc|cccccccc|}
\hline
 model  & $\dot m$ & $\mbh$\footnote{in unit of $\msun$} & $\zeta$ \hspace{5pt} 
 & \hspace{5pt} $\theta_e$ & $\tau_{\rm T}$ & $L_X$\footnote{X-ray luminosity in the 2-10 keV band in unit of $\rm erg~s^{-1}$} & $\gammapeq$ & $P_{\rm cr}/P_{\rm th}$ & $L_{\nu,\rm tot}$\footnote{Total neutrino luminosity in unit of $\rm erg~s^{-1}$}& $f_\pi$ & $\ecut$\footnote{in unit of GeV}  \\
\hline
A1 (reference) &  $1\times 10^{-2}$ & $10^7$ & 0.1 \hspace{5pt} & \hspace{5pt} 2.0 & $2.2\times10^{-2}$ &$3.0\times10^{39}$ & $1.4\times 10^5$ & 0.21 &$7.8\times10^{38}$& $2.2\times10^{-2}$ &  14   \\
 \hline
A2  & $1\times 10^{-3}$ & $10^7$ & 0.1 \hspace{5pt} & \hspace{5pt} 4.1 & $2.2\times10^{-3}$ & $2.6\times10^{38}$& $4.4\times10^4$ &0.20 & $6.4\times10^{36}$& $2.2\times10^{-3}$ &  24    \\
 \hline
A3  & $1\times 10^{-2}$ & $10^8$ & 0.1 \hspace{5pt} & \hspace{5pt} 2.1 & $2.2\times10^{-2}$ & $3.5\times10^{40}$&$4.4\times10^5$ &0.21 & $9.8\times10^{39}$ & $2.3\times10^{-2}$ &  5.4  \\
 \hline
A4  & $1\times 10^{-2}$ & $10^7$ & 0.3 \hspace{5pt} & \hspace{5pt} 2.0 & $2.2\times10^{-2}$& $3.0\times10^{39}$& $3.8\times10^6$& 0.21 & $3.1\times10^{39}$ & $7.7\times10^{-2}$ &  14   \\
 \hline
A5  & $6\times 10^{-2}$ & $10^7$ & 0.1 \hspace{5pt} & \hspace{5pt} 0.99 & $1.3\times10^{-1}$& $2.3\times10^{40}$ &$3.4\times10^5$ & 0.21 & $3.1\times10^{40}$ & $1.4\times10^{-1}$ &  1.3  \\
 \hline
\end{tabular}
\end{center}
\end{table*}

\section{Spectra of Non-thermal Particles in a Typical RIAF}\label{sec:spectra}

\subsection{Plasma in accretion flows}\label{sec:plasma}

If the infall time $\tfall$ is shorter than the relaxation time due to the Coulomb scattering $t_{\rm rel}$,
it allows the existence of non-thermal particles. The infall time for RIAFs is estimated to be
\begin{equation}
 \tfall \simeq \frac {R}{v_r} \sim 4.4 \times 10^4 r_1^{3/2} \alpha_{-1}^{-1} \mbhn ~ {\rm s}~, \label{eq:fall}
\end{equation}
whereas the proton-proton relaxation time is estimated as 
\begin{eqnarray}
  t_{\rm rel} &=& \frac{4\sqrt \pi} {\ln \Lambda} \frac{1}{n_p \sigmat c}
   \left(\frac{m_p}{m_e}\right)^2 \left(\frac{\kb T_p}{m_p c^2}\right)^{3/2} \nonumber\\
 &\sim& 2.1\times 10^7 \alpha_{-1} \mbhn \dot m_{-2}^{-1} \rm~ s
\end{eqnarray}
where $\ln \Lambda$ is the Coulomb logarithm \citep[e.g.,][]{spi62}. Thus, RIAFs satisfy $t_{\rm rel} \gg \tfall$, which allows $F(p)$ to be non-thermal \citep[cf.][]{tk85,mq97}. For RIAFs, $\tfall$ has the same order as the dissipation time via the $\alpha$ viscosity $t_{\rm dis}$ \citep[e.g.,][]{pri81}. Thus, the proton distribution function in RIAFs may not be Maxwellian within the dissipation time.

The protons inside RIAFs are scattered by turbulent magnetic fields. This process changes a momentum of each proton whose distribution function may be different from Maxwellian.
In this paper, we consider relativistic protons accelerated through stochastic acceleration in RIAFs.
It is expected that the stochastic acceleration leads to a hard spectrum of protons
with $s_p < 1$, where $dN_p/dE_p\propto E_p^{-s_p}$ \citep[e.g.,][]{bld06,sp08}. 
Thus, most of the accelerated protons accumulate on the high-energy end of proton distribution (see Equation (\ref{eq:epeq})). 
This implies that it is impossible to accelerate all the protons in RIAFs
because the protons are accelerated using the gravitational energy released by accretion, which is typically 0.1 $m_p c^2$ per a proton.
We assume only a small fraction of protons are injected to relativistic energy through some plasma processes, such as the magnetic reconnection \citep{hos13,hos15},
and those relativistic protons are governed by the Fokker-Plank equation \citep[e.g.,][]{sp08},
\begin{eqnarray}
 \delt F(p) &=& \frac{1}{p^2}\delp\left[p^2 \left(D_p \delp F(p) + \frac{p}{\tcool }F(p)\right) \right]\nonumber\\ 
&-& F(p)\left(\tdiff^{-1}+\tfall^{-1}\right)+\finj, \label{eq:fp}
\end{eqnarray}
where $F(p)$ is the distribution function of the non-thermal protons 
($dN_p/dE_p = 4\pi E_p^2 F(p) c $)
, $p$ is the momentum of the protons, $D_p$ is the diffusion coefficient for the momentum space, $\finj$ is the injection term, $\tcool$ is the cooling time, $\tdiff$ is the diffusion time, and $\tfall$ is the infall time.

When we consider the relativistic particles, we should compare the Coulomb loss time for relativistic particles $t_{\rm Coul}$ to $t_{\rm fall}$. The Coulomb loss time is estimated to be \citep[e.g.,][]{dml96}
\begin{eqnarray}
t_{\rm Coul}\simeq 1225\frac{(\gamma_p-1)(3.8\theta_e^{3/2}+1.0)}{\tau_{\rm T} \ln \Lambda} \frac R c \nonumber\\
 \sim 7\times10^7 r_1^{3/2} \alpha_{-1} \mbhn \dot m_{-2}^{-1} \theta_e^{3/2} \gamma_{p,1}\rm~ s
\end{eqnarray}
where $\gamma_p$ is the Lorentz factor of the proton. 
Since $t_{\rm Coul} > t_{\rm fall}$ is satisfied for RIAFs, we can neglect the Coulomb loss in RIAFs. 

It is considered that quasars have standard disks, in which the physical quantities are much different from those in RIAFs. For the Shakura-Sunyaev disks in the gas pressure dominant regime \citep[gas-SSD,][]{ss73}, we have longer $\tfall$ ($\tfall = R/v_r \simeq R/(\alpha \vk)(R/H)^2\sim 3\times10^8$ sec), and shorter $t_{\rm rel}$ ($\sim 3\times 10^{-9} {\rm ~ sec} \ll t_{\rm dis}$) than those of RIAFs. The dissipation time $t_{\rm dis}$ is the same as that of RIAFs (see Equation [\ref{eq:fall}]). Thus, $t_{\rm rel} \ll t_{\rm dis} \ll \tfall$ is satisfied in gas-SSDs. The distribution function $F(p)$ is expected to be Maxwellian due to the efficient Coulomb scattering.
Even for the relativistic particles, the Coulomb loss time is much shorter than the dissipation time for $\gamma_p \lesssim 10^3$ because they have large optical depth $\tau_{\rm T}\sim 10^4$ 
\citep[for the value of $\tau_{\rm T}$, see Equation (2.16) of][]{ss73}.
Therefore, it seems difficult to accelerate the particles in gas-SSDs. For other solutions, such as standard disks in the radiation pressure dominant regime \citep{ss73} and magnetically arrested disks \citep{br74}, the Thomson optical depth may not be as large as gas-SSDs, and it might be possible to satisfy $t_{\rm dis} < t_{\rm Coul}$.

\subsection{Timescales}

Equation (\ref{eq:fp}) involves four important timescales, the acceleration time $\taccel\equiv p^2/D_p$, the diffusion time $\tdiff$, the infall time $\tfall$, and the cooling time $\tcool$. 

In this paper, we assume a power spectrum $P(k)\propto k^{-q}$, and fix the index of the power spectrum $q=5/3$ for simplicity. This value is motivated by the Alfv\'enic turbulence \citep{gs95}, although other modes may also play an important role on particle acceleration. According to the quasi-linear theorem, the diffusion coefficient is \citep[e.g.,][]{dml96}
\begin{equation}
 D_p \simeq  (m_p c)^2 (c \kmin) \left(\frac{\va}{c}\right)^2 \zeta (r_L \kmin)^{q-2} \gamma_p^q ,
\end{equation}
where $\kmin \sim R^{-1}$ is the minimum wave number of the turbulence, $\va= B/\sqrt{4\pi m_p n_p}$ is the Alfven speed, $r_L = m_p c^2/(eB)$, and $\zeta= 8\pi\int P(k)dk/B_0^2$ is the ratio of the strength of turbulent fields to that of the non-turbulent fields. Then, the acceleration time is
\begin{eqnarray}
 \taccel &\simeq& \frac{p^2}{D_p} \simeq \frac{1}{\zeta} \left(\frac{v_{\rm A}}{c}\right)^{-2} 
  \frac R c \left(\frac {r_L}{R}\right)^{2-q} \gamma_p^{2-q}\nonumber \\
  &\sim& 1.1\times 10^3 r_{1}^{25/12}\alpha_{-1}^{1/6} \beta_3^{7/6} \mbhn^{5/6} \dot m_{-2}^{-1/6} \zeta_{-1}^{-1}\nonumber\\
  &\times& \gamma_{p,1}^{1/3}\rm ~ s.   
\end{eqnarray}

We consider the diffusive escape and infall as the sink term of Equation (\ref{eq:fp}).
The particles fall to the SMBH in the infall time, given by Equation (\ref{eq:fall}).
For isotropically turbulent magnetic fields, the diffusion time is \citep[e.g.,][]{sp08}
\begin{eqnarray}
 \tdiff &\simeq& \frac {9R} {c} \zeta \left(\frac{r_L}{R}\right)^{q-2}\gamma_p^{q-2} \nonumber \\
   &\sim& 6.7\times 10^6 r_{1}^{11/12}\alpha_{-1}^{-1/6} \beta_3^{-1/6} \mbhn^{7/6} \dot m_{-2}^{1/6} \zeta_{-1}^{1} \nonumber\\
   &\times&\gamma_{p,1}^{-1/3} \rm ~ s.   
\end{eqnarray}

For the cooling time, we consider inelastic $pp$ and $p\gamma$ reactions, and the proton synchrotron emission process. The total cooling rate is given as 
\begin{equation}
 \tcool^{-1} = t_{pp}^{-1} + t_{p\gamma}^{-1} + \tsync^{-1},
\end{equation}
where $t_{pp}$, $t_{p\gamma}$, and $\tsync$ are cooling time scales for each process.  We neglect the inverse Compton scattering by protons and the Bethe-Heitler process because they are typically sub-dominant.  The synchrotron cooling rate is 
\begin{equation}
 \tsync^{-1} = \frac{4}{3} \left(\frac{m_e}{m_p}\right)^3\frac{c\sigmat U_B}{m_e c^2} \gamma_p, 
\end{equation}
where $U_B=B^2/(8\pi)$ is the energy density of the magnetic fields.
The $pp$ cooling rate is  
\begin{equation}
 t_{pp}^{-1} = n_p\sigma_{pp} c K_{pp},
\end{equation}
where $K_{pp}\sim 0.5$ is the proton inelasticity of the process. The total cross section of this process $\sigma_{pp}$ is represented as a function of the proton energy $E_p$, 
\begin{equation}
   \sigma_{pp} \simeq (34.3+1.88L + 0.25L^2)\left[1-\left(\frac{E_{pp,\rm thr}}{E_p}\right)^4\right]^2~ \rm mb
\end{equation}
for $E_p\ge E_{pp,\rm thr}$, where $L=\log(E_p/1\rm TeV)$ and $E_{pp,\rm thr}=$1.22 GeV \citep{kab06}.
The $p\gamma$ cooling rate is 
\begin{eqnarray}
 t_{p\gamma}^{-1} = \frac{c}{2\gamma_p^2}
  \int_{\bar{\varepsilon}_{\rm thr}}^{\infty}d\bar{\varepsilon}\sigma_{p\gamma}(\bar{\varepsilon})K_{p\gamma}(\bar{\varepsilon})\bar{\varepsilon} \nonumber \\
  \times \int_{\bar{\varepsilon}/(2\gamma_p)}^{\infty}dE_\gamma \frac{N_\gamma(E_\gamma)}{E_\gamma^2},
\end{eqnarray}
where $\bar{\varepsilon}$ and $E_\gamma$ are the photon energy in the proton rest frame and the black hole frame, respectively, 
$N_\gamma(E_\gamma)$ is the photon occupation number, and $\bar{\varepsilon}_{\rm thr} = $145 MeV. 
We use the rectangular approximation for this process \citep{ste68}.  Assuming $\sigma_{p\gamma}(\bar{\varepsilon})K_{p\gamma}(\bar{\varepsilon})=\delta(\bar{\varepsilon}-\epeak)\sigmapeak \Kpeak \Delta \epeak$, we write $t_{p\gamma}$ as 
\begin{eqnarray}
 t_{p\gamma}^{-1} = \frac{c}{2\gamma_p^2} \epeak \Delta \epeak \sigmapeak \Kpeak \nonumber \\
  \times \int_{\epeak/(2\gamma_p)}^{\infty}dE_\gamma\frac{N_\gamma(E_\gamma)}{E_\gamma^2}, 
\end{eqnarray}
where $\epeak\sim 0.3$ GeV, $\sigmapeak\sim 5\times 10^{-28}\rm ~cm^2$, $\Kpeak\sim 0.2 $, $\Delta \epeak \sim 0.2$ GeV.

  \begin{figure*}[t]
    \includegraphics[width=\linewidth]{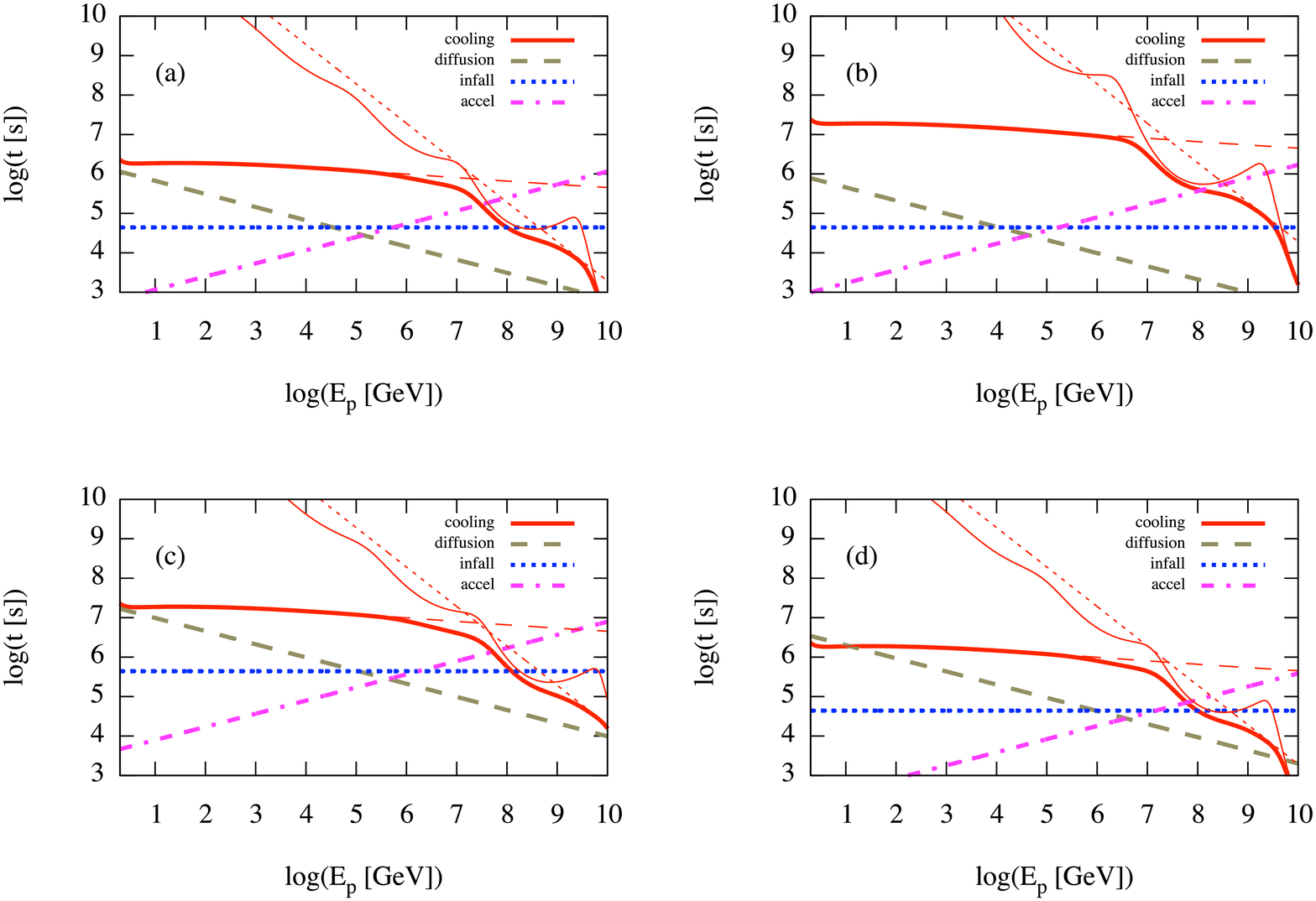}
   \centering
   \caption{Energy dependence of the timescales.  
   We plot the cooling time (thick-solid), the diffusion time (thick-dashed),
   the infall time (thick-dotted), and the acceleration time (dotted-dashed). 
   The thin-solid, thin-dashed, and thin-dotted lines show the $t_{p\gamma}$ ,$t_{pp}$, and $\tsync$, respectively. 
   Panels (a)--(d) show the cases for models A1 (reference), A2 (low $\dot m$), A3 (high $\mbh$), 
   and A4 (high $\zeta$), respectively. 
   \label{fig:times}}
   \end{figure*}

Figure \ref{fig:times} shows the timescales for models A1, A2, A3, and A4, whose parameters are tabulated in Table \ref{tab:single}. For low values of $E_p$, $\taccel$ is the shortest for all the models. At some energy $\epeq$,
$\taccel=\tdiff$ is satisfied.  Above the energy, the acceleration is limited by the diffusive escape. Equating $\tdiff$ and $\taccel$, we can estimate $\gammapeq \equiv \epeq/(m_p c^2)$ to be
\begin{eqnarray}
 \gammapeq
 &\sim&  \left(\frac{3 \zeta v_{\rm A}}{c}\right)^3\left(\frac{R}{r_L}\right)\nonumber \\
 &\sim& 1.4\times 10^{5} \dot m_{-2}^{1/2}\mbhn^{1/2}\alpha_{-1}^{1/2}\zeta_{-1}^{3}\beta_3^{-2} r_1^{-7/4}. \label{eq:epeq}
\end{eqnarray}
We tabulate the value of $\gammapeq$ in Table \ref{tab:single}. 
This characteristic energy strongly depends on $\zeta$. The higher $\dot m$ or lower $\beta$ makes the magnetic fields stronger, so that the $\gammapeq$ is higher. The larger $r$ weakens $B$, which leads to lower $\gammapeq$.
The estimation in Equation (\ref{eq:epeq})  is correct as long as we choose $\beta \lesssim 10$.
As $\beta$ becomes higher, $\taccel$ becomes longer but $\tfall$ does not change.
Thus, the infall limits the acceleration for $\beta \gtrsim 10$.
We note that $\epeq$ does not correspond to the peak energy of the $E_p L_{E_p}$ spectrum (see the next subsection). When diffusive escape limits acceleration, the distribution function declines gradually above $\epeq$,
whose asymptote is $F(p) \propto E_p^{-3/2} \exp(-(27E_p/\epeq)^{1/3})$ for $q = 5/3$ \citep[see Equation (56) of][]{bld06}. 
This allows the protons to have about 10 times higher energy than the estimate in Equation (\ref{eq:epeq}). Thus, LLAGN can have the protons up to $E_p\gtrsim 10^{16}$ eV when $\zeta \gtrsim 0.2$.

For all the models, at low energies, $pp$ inelastic collisions dominate over the synchrotron and photomeson production processes.
At high energies around $E_p \gtrsim 10^6 - 10^7$ GeV, 
the photomeson production becomes relevant although the synchrotron cooling is comparable to it.
For large $\beta$ or large $\delta_e$ cases, the $p\gamma$ reaction is more efficient than the synchrotron
owing to the high target photon density. 

\subsection{Spectra of Non-thermal Particles}

When we solve Equation (\ref{eq:fp}), we treat the injection term as a delta-function $\finj = F_0\delta(p-\pinj)$, where $\pinj$ is the injection proton momentum and $F_0$ is the normalization factor of injection. We fix $\pinj = 2 m_p c $  because $\pinj$ little affects the profile of distribution function as long as we choose $\pinj c \ll \epeq$.
We assume that the total luminosity expended to inject and accelerate relativistic protons is proportional to the accretion luminosity, $\dot M c^2$.
As seen in the previous subsection, the proton acceleration is limited by escape. 
We determine the normalization of relativistic protons
such that the luminosity of injection and acceleration balances with the escape luminosity, i.e., 
\begin{equation}
\etacr\dot M c^2 = \int dV \int dp 4\pi p^2 F(p) E_p \left(\tfall^{-1}+\tdiff^{-1}\right),
\end{equation}
where $\etacr$ is a parameter of injection efficiency.
This parameter determines the normalization of the non-thermal protons, not affecting the shapes of the spectra. 
\citet{ktt14} shows that the non-thermal particles do not substantially affect the dynamical structure if $\etacr \lesssim 0.1$. 
We use $\etacr = 0.01$ as a fiducial value. 

We solve Equation (\ref{eq:fp}) until steady solutions are realized by using the Chang-Cooper method \citep{cc70}. We set the computational region from $E_p = 1.5$ to $10^{10}$ GeV  and divide the grids so that they are uniform in the logarithmic space. The number of the grid points is $N=500$. We calculate some models with $N=1000$ and find that the results are unchanged by the number of grids. 

From the calculation results, we estimate the cosmic-ray pressure defined as 
\begin{equation}
 P_{\rm cr} = 4\pi \int dp  p^2 f \frac{cp}{3}. 
\end{equation}
We tabulate the ratio of cosmic-ray pressure to thermal pressure in Table \ref{tab:single}. 
We find that $P_{\rm cr}\propto \etacr$, and $P_{\rm cr}/P_{\rm th}$ is almost independent of the other parameters. 
For all the models, $P_{\rm cr} < P_{\rm th}$ is satisfied,
and thus, it is justified that non-thermal protons do not affect the dynamical structure of RIAFs \citep{ktt14}.

\begin{figure}
   \centering
    \includegraphics[width=\linewidth]{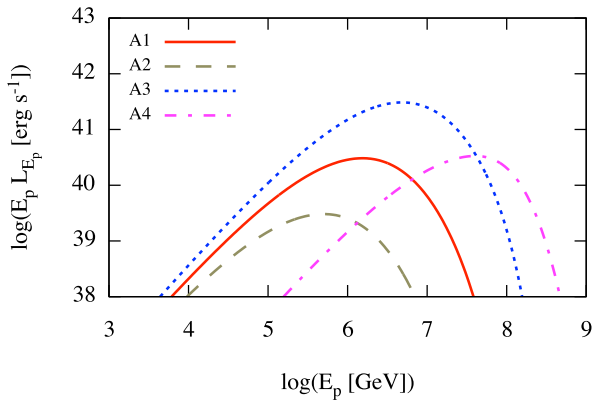}
   \caption{Differential luminosity spectra of escaping protons
   for models A1 (red-solid), A2 (green-dashed), A3 (blue-dotted), and A4 (magenta dotted-dashed), respectively. 
   \label{fig:escpro}}
   \end{figure}

Since the peak energy is determined by CR escape for all the models, the profiles of the distribution functions are quite similar to each other. They show a power law $F(p) \propto E_p^{-(1+q)}$ for low $E_p$
\citep[see Equation 56 of][]{bld06}. For $E_p \gtrsim \epeq$, they deviate from the power-law and their asymptote is $F(p) \propto E_p^{-3/2} \exp(-(27E_p/\epeq)^{1/3})$ for $q = 5/3$ \citep{bld06}.
After obtaining $F(p)$, we estimate the differential luminosity spectra of the escaping protons to be
\begin{equation}
 E_p L_{E_p} = \int dV \frac{4\pi p^3 F(p) E_p}{\tdiff} = \frac{4 \pi^2 c R^3 p^4 F(p)}{\tdiff}.
\end{equation}
We plot $E_p L_{E_p}$ in Figure \ref{fig:escpro}. We tabulate parameter sets in Table \ref{tab:single}, fixing the parameters $\alpha=0.1$, $\beta=3$, $r=10$, $\delta_e=3\times10^{-3}$, $q=5/3$, and $\etacr=0.01$.
For A1, the total luminosity of protons is $\sim 1\times10^{41}\rm~erg~s^{-1}$, and the differential luminosity has a peak $E_p L_{E_p}\sim 3\times10^{40} \rm~erg~s^{-1}$ at $E_p \sim 1.5$ PeV. 
Since the total luminosity of escaping protons is proportional to the released energy $\dot M c^2$, the peak luminosity of escaping protons is almost proportional to $\dot m$ and $\mbh$ (see A2 and A3 in Figure \ref{fig:escpro}), while it is almost independent of other parameters. 
All the models have a power law, $E_p L_{E_p} \propto E_p ^{5-2q}$, for $E_p < \epeq$.
For $E_p > \epeq$, the spectra deviate from the power law
and their asymptote is $E_p L_{E_p}\propto E_p^{17/6}\exp(-(27E_p/\epeq)^{1/3})$ for $q=5/3$ \citep[see Equation (70) of][]{bld06}.
This makes a peak at the energy $\epmax \sim 10 \epeq$. 
The parameter dependence of $\epmax$ is consistent with the estimation by Equation (\ref{eq:epeq}).

The neutrino spectrum is estimated to be
\begin{equation}
 E_\nu L_{E_\nu} = \left(\frac{1}{2t_{pp}}  +\frac{3}{8t_{p\gamma}} \right) 4\pi^2 R^3 p^3 E_p F(p), 
\end{equation}
where $E_\nu = 0.05 E_p$ is the neutrino energy.  As long as the $pp$ reaction is the dominant process of neutrino production, this treatment becomes invalid for spectra that are harder than $F(p) \propto  p^{-2.5} - p^{-2.7}$ \citep[e.g.,][]{kab06}.  Since we expect hard proton spectra $F(p) \propto p^{-(1+q)}$ with $q=5/3$, our analytical method to calculate neutrino spectra will not be accurate at low energies. Thus, we show neutrino spectra only at $E_\nu > 1$ TeV energies.

\begin{figure}
   \centering
    \includegraphics[width=\linewidth]{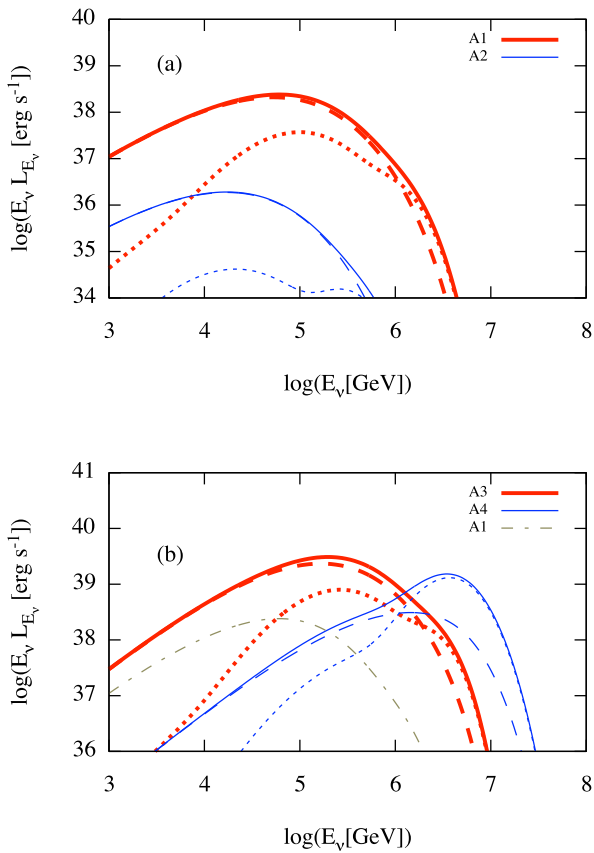}
   \caption{Differential luminosity spectra of neutrinos.
    The solid lines represent the total neutrino spectra.
    The dashed and dotted lines show the neutrino spectra from $pp$ and $p\gamma$ interactions, respectively.
    Panel (a) shows the models A1 (red-thick lines) and A2 (blue-thin lines). 
    Panel (b) shows the models A3 (red-thick lines) and A4 (blue-thin lines).
    The thin dotted-dashed line (total spectra for A1) is also plotted for comparison in Panel (b). 
   \label{fig:escneu}}
   \end{figure}

Figure \ref{fig:escneu} depicts spectra of neutrinos, 
The neutrinos are mainly made via the $pp$ collisions for A1, A2, A3, because $t_{pp} < t_{p\gamma}$ for $E_p \lesssim \epmax$.  The $pp$ cooling rate is almost independent of the proton energy.  Thus, neutrino spectra are similar to those of protons unless proton spectra are too hard. The neutrino luminosity at the peak is estimated to be $E_\nu L_{E_\nu}|_{\enumax} \propto \etacr \dot m^2 \mbh \alpha^{-1}\beta^{1/2}$, where $\enumax = 0.05\epmax$ is the peak neutrino energy. 
We can see this feature in Figure \ref{fig:escneu} by comparing the dashed lines.
On the other hand, both the $pp$ and $p\gamma$ processes are important for A4. 
The photomeson production is dominant for $E_\nu \gtrsim 10^6$ GeV. 
This makes another peak in the spectra
because the neutrino spectrum by $p\gamma$ reactions reflects the target photon spectrum. 
For example, in A4, the target photon field has a bump made by the inverse Compton scattering at $E_\gamma \sim 2$ eV, which leads to a peak in the neutrino spectrum at $E_\nu \sim 3 \times 10^{6}$ GeV.  
The total neutrino luminosity, $L_{\nu,\rm tot}$, is tabulated in Table \ref{tab:single}. 

The gamma rays are also emitted through pion decay. The total luminosity of the gamma rays would be the same order as the neutrino luminosity, although the spectrum is different because of the pair production process (see Section \ref{sec:gamma}).

Since proton acceleration is limited by escape in our models, the total injection luminosity is almost the same as the proton escape luminosity. 
This allows us to write the efficiency of pion production as
\begin{equation}
 f_\pi \approx \frac{t_{pp}^{-1} + t_{p\gamma}^{-1}}{\tfall^{-1}+ \tdiff^{-1}}. 
\end{equation}
If $t_{pp} < t_{p\gamma}$ and $\tdiff < \tfall$ at $\epeq$,
we can write the parameter dependence of pion production efficiency at $\epeq$ as $f_\pi \propto \dot m \alpha^{-1} \beta^{1/2}$.  Thus, LLAGN with high $\dot m$ can emit neutrinos more efficiently than those with low $\dot m$.
On the other hand, we cannot simply write down the parameter dependence of neutrino luminosity for $p\gamma$ dominant cases because it depends on the target photon spectrum.  The efficiency of pion production is 
tabulated in Table \ref{tab:single}. For all models, the pion production efficiency is $f_\pi \lesssim 0.1$. 
Thus,
most of the high-energy protons escape from RIAFs without losing their energies. 

If we consider models that have high $\delta_e$,
$\zeta$, and $\dot m$, compared to A1, CR acceleration is limited by the photomeson production because high $\zeta$ increases $\tdiff$, and high $\dot m$ and/or $\delta_e$ decrease $t_{p\gamma}$. 
In this case, the scaling of the peak energy is different from the one obtained with Equation (\ref{eq:epeq}), and proton spectra could change \citep[see, e.g.,][]{sp08}.  However, this parameter range looks extreme in our model.  High $\dot m$ and $\delta_e$ lead to high photon luminosities, which are inconsistent with the concept of RIAFs.  In addition, $\zeta$ should be less than unity for the validity of the quasi-linear theory.  Thus, we can focus on the models where the diffusive escape limits the acceleration.  

\section{Diffuse Intensities of Neutrinos and Cosmic-Ray Protons} \label{sec:diffuse}

The diffuse neutrino intensity from extragalactic sources is given by \citep[e.g.,][]{am04,mid14}
\begin{eqnarray}
 \Phi_\nu = \frac{c}{4\pi H_0}\int_0^{\zmax} \frac{dz}{\sqrt{\Omega_M(1+z)^3 + \Omega_{\Lambda}}}\nonumber \\
\times \int_{\lmin}^{\lmax} d L_{\rm bol} \phi(L_{\rm bol},~z) \frac{L_{E_\nu'}(L_{\rm bol})}{E_\nu'}, \label{eq:diffuse}
\end{eqnarray}
where $\phi(L_{\rm bol},z)$ is the luminosity function, $E_\nu' = (1+z)E_\nu$ is the neutrino energy at the rest flame of LLAGN.
We assume that LLAGN exist from $z=0$ to $z=\zmax$ and from $\lmin$ to $\lmax$. 

The luminosity function of $H_\alpha$ from nearby LLAGN is plotted in Figure 8 of \citet{ho08}.  
Here we assume a broken power law shape,
\begin{equation}
 \phi_0(L_{H_\alpha})=\frac{n_\ast/ L_\ast}{\left(L_{H_\alpha}/L_\ast\right)^{s_1} + \left(L_{H_\alpha}/L_\ast\right)^{s_2}}. 
\end{equation}
From Figure 8 of \citet{ho08}, we find $L_\ast=10^{38}\rm ~erg~s^{-1}$, $n_\ast\sim 1.3\times 10^{-2}\rm Mpc^{-3}$, $s_1\sim1.64$, $s_2\sim1$ between $3\times10^{36}~{\rm erg~s^{-1}}  < L_{H_\alpha} < 3\times10^{41}~{\rm erg~s^{-1}}$.
For LLAGN, $L_{H_\alpha}$ is related to the bolometric luminosity $L_{\rm bol}$ as $L_{\rm bol}\sim 80 L_{H_\alpha}$ \citep{ho08}.
Since the redshift evolution is poorly known, we assume no evolution of the luminosity function $\varphi(L_{\rm bol},z)=\varphi_0(L_{\rm bol})$.  This is because LLAGN are similar to the BL Lac objects in the sense that they have a faint disk component. The luminosity function of BL Lac objects is nearly consistent with no evolution \citep{aje+14}.

\begin{figure}
   \centering
    \includegraphics[width=\linewidth]{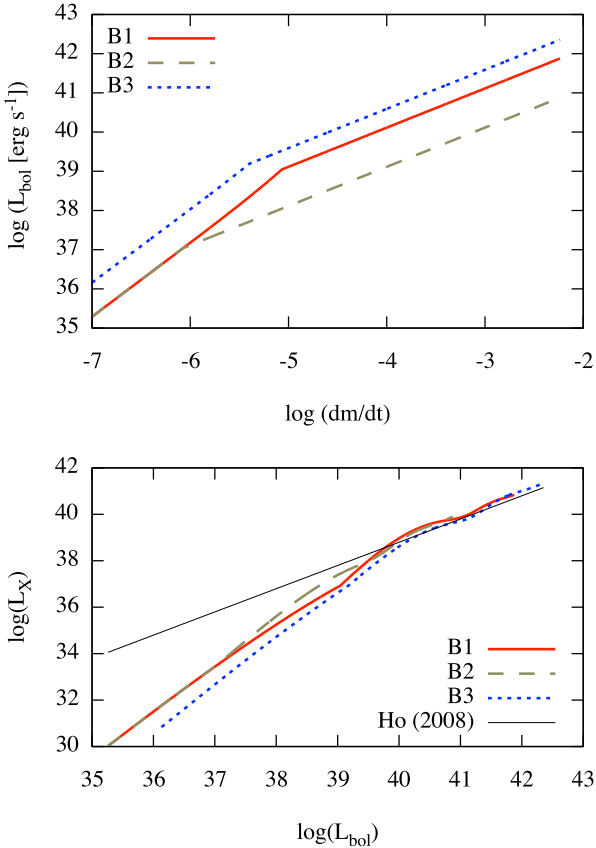}
   \caption{The bolometric and X-ray luminosities in our models. 
    The top panel shows $L_{\rm bol}$ as a function of $\dot m$ for each model tabulated in Table \ref{tab:diffuse}.
    The bottom panel shows $L_X$ as a function of $L_{\rm bol}$. 
    The thick lines are the resultant values for our models. 
    The thin-solid line in bottom panel is the averaged value based on observations \citep{ho08}. 
    The lines for model B4 are not shown, since they are completely the same as those for model B3.
   \label{fig:luminosity}}
   \end{figure}

In our RIAF model, we can calculate $L_{\rm bol}$ for given $\mbh$, $\dot m$, $\delta_e$, and $\beta$ as described in Section \ref{sec:targetphoton}. We assume that all the LLAGN have the same values of $\mbh$, $\delta_e$, $\beta$. Then, we can integrate Equation (\ref{eq:diffuse}), using the relationship of $\dot m$ to $L_{\rm bol}$, where $L_{\rm bol}$ is a monotonically increasing function of $\dot m$ as shown in the upper panel of Figure \ref{fig:luminosity}. The parameters are tabulated in Table \ref{tab:diffuse}. 
The break of each line corresponds to the change of $\theta_e$ from $\theta_{e,\rm vir}$ to $\theta_{e,\rm eq}$. 
In reality, other components such as the radiation from jets may contribute to $L_{\rm bol}$ from LLAGN, but we ignore the other components for simplicity.
The bottom panel shows the X-ray luminosity in the 2-10 keV band, $L_X$, 
as a function of $L_{\rm bol}$. We can see that the deviation between our model and observation is not large in $L_{\rm bol} \gtrsim 10^{39} \rm ~erg~s^{-1}$, although the faint part is quite different. 
We set $\mdmin=10^{-7}$ but the detailed value of $\mdmin$ little affects the results. 
A RIAF is expected to change to a standard thin disk above the maximum accretion rate $\mdmax$. 
We use $\mdmax = 0.06 $ following to \citet{xy12}.
We tabulated the physical quantities for LLAGN with $\mdmax$ as model A5 in Table \ref{tab:single}. 
Then, the maximum luminosity is written as
\begin{equation}
\lmax= \delta_e \mdedd \mdmax c^2 \simeq 7.6 \times 10^{41} \mbhn \delta_{e,-2} \rm ~erg~s^{-1}.
\end{equation}
The maximum luminosity for each model is tabulated in Table \ref{tab:diffuse}. 
We calculate the diffuse spectra with some values of $\zmax$ and confirm that $\zmax$ does not affect the results if we use a sufficiently high value $\zmax\gtrsim 4$.  
Recent studies show that the number density of $\mbh$ is high at $\mbh\sim 10^7 M_{\odot}$ and monotonically decreases with $\mbh$ \citep[e.g.,][]{lhw11}.  On the other hand, it seems that the average of $\mbh$ of nearby LLAGN
whose multi-band spectra are observed is $\sim10^8 M_{\odot}$ \citep{ehf10}.
This suggests that the average mass of SMBHs in LLAGN is not so clear. We calculate two cases for $\mbh=10^7 M_{\odot}$ and $\mbh=10^8 M_{\odot}$.
Fixing $q=5/3$, $\alpha=0.1$, $\beta=3$, and $r=10$,
we search suitable $\zeta$,  $\delta_e$, and $\etacr$ 
to see whether the calculated intensity amounts to the observed one. 
\begin{table*}[tb]
\begin{center}
\caption{Models for diffuse neutrino flux: parameters and resultant quantities \label{tab:diffuse}}
\begin{tabular}{|c|cccc|ccccc|}
\hline
 model & $\mbh$\footnote{in unit of $\msun$} & $\delta_e$ & $\zeta$ & $\etacr$  \hspace{10pt}
 &  \hspace{10pt}  $f_{\pi,\rm max}$ & $L_{\rm max}$ $^{\rm b}$ & $L_{\rm mid}$ $^{\rm b}$
 & $L_{\rm max,X}$ $^{\rm b}$ & $L_{\rm mid,X}$\footnote{in unit of erg s$^{-1}$}\\
\hline
B1 & $10^7$ & 0.01 & 0.18 & $6.0\times 10^{-3}$ \hspace{10pt} &  
		     \hspace{10pt} 0.19 & $7.6\times10^{41}$ & $7.8\times10^{40}$ & $6.7\times10^{40}$ & $6.3\times10^{39}$ \\
 \hline
B2 & $10^7$ & 0.001 & 0.06 & $2.0\times 10^{-2}$ \hspace{10pt} &
		     \hspace{10pt} 0.13 & $7.6\times10^{40}$ & $2.5\times10^{40}$ & $7.8\times10^{39}$ & $2.6\times10^{39}$ \\
 \hline
B3 & $10^8$ & 0.003 & 0.14 & $1.4\times 10^{-3}$ \hspace{10pt} &
		     \hspace{10pt} 0.15 & $2.3\times10^{42}$ & $1.4\times10^{41}$ & $2.2\times10^{41}$ & $6.9\times10^{39}$ \\
 \hline
B4 & $10^8$ & 0.003 & 0.04 & $1.5\times 10^{-2}$ \hspace{10pt} &  
		     \hspace{10pt} 0.15 & $2.3\times10^{42}$ & $1.4\times10^{41}$ & $2.2\times10^{41}$ & $6.9\times10^{39}$ \\
 \hline
\end{tabular}
\end{center}
\end{table*}

\subsection{Diffuse intensity of neutrinos}

In this subsection, we show that our models can fit spectra of neutrinos observed by IceCube \citep{ice14}. 
Recently, IceCube reported neutrino spectra around 10 TeV \citep{ice15}.  The flux around 10 TeV is higher than that at PeV energies.  Although it may be premature to discuss the origin of this low-energy excess, we show that it is possible for LLAGN to explain the excess. 

\begin{figure}
   \centering
    \includegraphics[width=\linewidth]{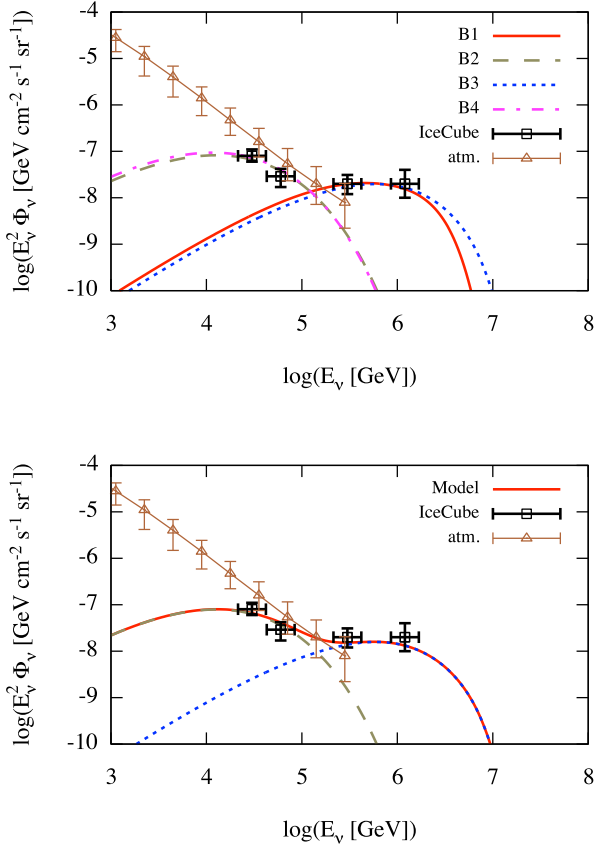}
   \caption{The diffuse neutrino intensity (per flavor) from RIAFs in the LLAGN model. 
    The top panel shows the diffuse neutrino intensity for each model tabulated in Table \ref{tab:diffuse}.
    The dashed line (B2) almost overlaps the dotted-dashed line (B4). 
    The bottom panel shows the diffuse intensity from two-component model (see text for detail).
    The red-solid, green-dashed, and blue-dotted lines show the total intensity, intensity from low-energy part,
    and intensity from high-energy part, respectively. 
    The green triangles represent the atmospheric muon neutrino background produced by CRs. 
    The black squares show the observed data of neutrino signals.
   \label{fig:diffuse}}
   \end{figure}

The results are plotted in upper panel of Figure \ref{fig:diffuse}, 
whose parameter sets are tabulated in Table \ref{tab:diffuse}.
LLAGN can emit enough neutrinos to amount to the observed flux with reasonable parameter sets.
In view of the PeV neutrino observation, protons must be accelerated up to around several tens of PeV energies, and their spectra cannot be extended to higher energies.  This is feasible unless $\zeta$ is somehow very low.
The spectral shape is affected by $\delta_e$ and $\zeta$ because these parameters
determine whether the photomeson production is important or not.
The injection efficiency $\etacr$ just determines the normalization of the diffuse neutrino flux, and
we find that $\etacr \sim 10^{-3}-10^{-2}$ to account for the PeV neutrinos.
The spectra for B1 and B3 can fit the data at 0.1--1 PeV energies, although we have difficulty in explaining the 10--100 TeV neutrino flux at the same time. They are almost flat for
0.1 PeV $\lesssim E_\nu \lesssim$ 1 PeV and have a cutoff at a few PeV energies.
We can also fit the data of 10 TeV neutrinos with lower values of $\zeta$ (B2 and B4).
They have a peak at $E_\nu \sim 10$ TeV and gradually decrease for $E_\nu > 10$ TeV.
The photomeson production is ineffective in these models because of the lack of target photons. 
Although we need a higher $\etacr$ for 10 TeV neutrinos, 
the required injection efficiency $\etacr \sim 0.01$ is lower
than that for other AGN models \cite[cf.,][]{am04,mid14}. 
Even if each LLAGN is much fainter than quasars, the number density of LLAGN is so high that they can significantly contribute to the diffuse neutrino flux in principle.

We show four models in the upper panel of Figure \ref{fig:diffuse} for demonstration,
but it is possible to fit the data with other sets of parameters. 
For the models with higher $\delta_e$,
higher $\etacr$ is needed to achieve the observed flux. 
This is because higher $\delta_e$ makes $\lmax$ higher, and thereby the corresponding number density of LLAGN decreases.
The higher $\delta_e$ also causes more efficient $p\gamma$ reaction. 
Thus, the maximum $f_\pi$ for model B1 has 0.19,
 which is larger than other models (see Table \ref{tab:diffuse}).
When we consider higher $\beta$, the fitting requires higher $\zeta$.
For example, the models with $\beta=10$ and $\zeta=0.3$ can fit the PeV neutrino data. 
Note that even for $\zeta\sim 1$,
we can use the quasi-linear theory for describing the stochastic acceleration
if $v_{\rm A}/c\gtrsim 0.1$ \citep[e.g.,][]{ort09}.
Since $v_{\rm A}/c\sim 0.1 r_1^{-1/2}\beta_3^{-1/2}$ for our model,
it is acceptable to use $\zeta\lesssim 0.3$ for the fitting of PeV neutrino data. 

Since each LLAGN model cannot fit both 10--100 TeV and 0.1--1 PeV data simultaneously,
it is natural to consider that either of them originates from other sources, 
such as low-luminosity gamma-ray bursts \citep{mi13} and starburst galaxies \citep{tam14}. 
If we tentatively consider a two-component model of LLAGN, it is possible to explain all the data. 
For example, suppose that 80\% of LLAGN have the parameters of model B2 except for $\etacr=2.4\times10^{-2}$.
The others have a parameter set of model B3 except for $\etacr= 5.6\times10^{-3}$.
The bottom panel of Figure \ref{fig:diffuse} shows the result of this two-component model. 
The resulting spectrum could fit all the four data points at the same time.
Although this two-component model is a possibility to explain the IceCube events,
this parameter choice is ad hoc.
The combination of LLAGN model with other sources would be more natural.
For example, the low-energy part could come from LLAGN,
while the high-energy part may come from CR reservoirs such as starburst galaxies and galaxy clusters. 

\begin{figure}
   \centering
    \includegraphics[width=\linewidth]{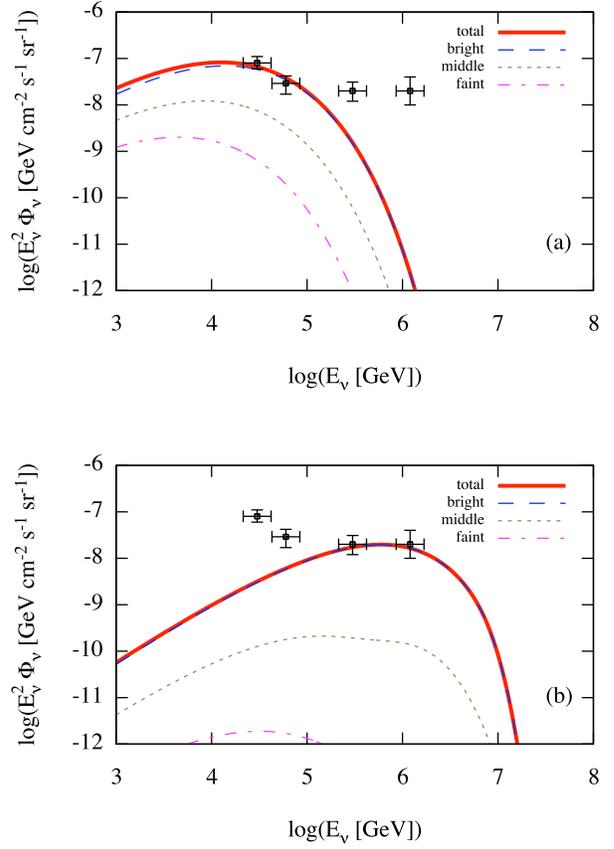}
   \caption{The contribution to the total intensity (red-thick lines) from different luminosity bins (thin lines).
    The blue-dashed, green-dotted, and magenta dotted-dashed lines 
    show the fluxes from bright, middle, and faint parts, respectively.
    See text for definition of the each part.
    The black squares show the observed data of neutrino signals.
    The top and bottom panels show the intensity for B2 and B3, respectively. 
   \label{fig:lx_depend}}
   \end{figure}

The diffuse neutrino flux is dominated by LLAGN with high $\dot m$ in our model.
The neutrino luminosity is higher as $\dot m$ is higher,
while the number density of LLAGN is lower for higher $\dot m$.
The former is more efficient than the latter for the neutrino luminosity.
We show the contribution to the total intensity from different luminosity bins in Figure \ref{fig:lx_depend}.
We set the luminosity bins as a faint part $L_{\rm min}<L_{\rm bol}< 80L_\ast$,
a middle part $80L_\ast < L_{\rm bol} < L_{\rm mid}$, and a bright part $L_{\rm mid} < L_{\rm bol} < L_{\rm max}$,
where $L_{\rm mid} = \sqrt{80 L_\ast L_{\rm max}}$ whose values are tabulated in Table \ref{tab:diffuse}.
We also tabulate the corresponding values of $L_X$ to $L_{\rm max}$ and $L_{\rm mid}$. 
The bright part emits almost all the neutrinos for all models. 
The faint and middle parts little contribute to the diffuse neutrino flux due to the low pion production efficiency.


\subsection{Diffuse intensity of cosmic-ray protons}

In our model, most of the injected protons escape from the accretion flow
without depletion due to the low efficiency of pion production $f_\pi\lesssim0.2$.
Here, we discuss the effects of escaping protons. 

\begin{figure}
   \centering
    \includegraphics[width=\linewidth]{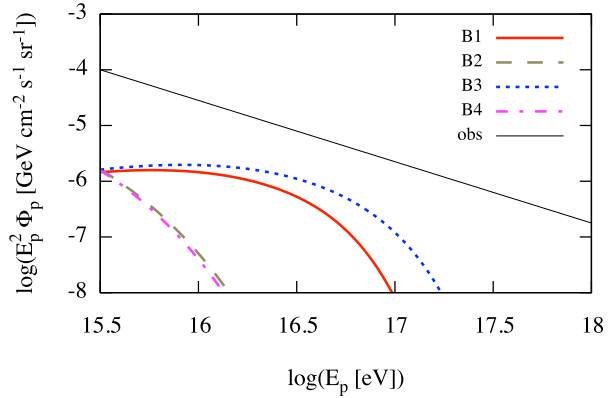}
   \caption{The maximum flux of the diffuse CR protons.
    The thick lines  show the CR flux
    for B1 (red-solid), B2 (green-dashed), B3 (blue-dotted), and B4 (magenta dotted-dashed).
    The thin-black-solid line shows the observed CR flux \citep[e.g.,][]{bec08}.
    The dashed line (B2) almost overlaps the dot-dashed line (B4).
   \label{fig:cr}}
   \end{figure}

Assuming that the Universe is filled with CR protons,
we can estimate the CR flux as in the neutrino flux.  
Figure \ref{fig:cr} shows the estimated flux of CR protons for models B1, B2, B3, and B4.
This flux of the escaping protons is much lower 
than observed CR flux for $10^{15.5} {\rm eV} < E_p < 10^{18}$ eV for all the models.
Although the escaping proton luminosity has weaker dependence on $\dot m$ than that of neutrino luminosity,
the bright part is dominant for the CR proton flux.

We note that it is unclear whether CRs of $E_p \sim 10^{16}$ eV are able to arrive at the Earth from LLAGN. 
In fact, the magnetic fields in the intergalactic medium (IGM) prevent the protons from traveling straightly, 
so that the distant sources cannot contribute to the CR flux. 
The diffusion length of CR protons during the cosmic time is estimated to be $\sim 6 B_{-8}^{-1/6} E_{p,16}^{1/6} l_{\rm coh,2}^{1/3}$ Mpc ($E_p\lesssim 10^{18}$ eV),
where we use $B_{-8}=B/(10^{-8}{\rm~Gauss})$, $E_{p,16}=E_p/(10~\rm PeV)$, 
and the coherence length $l_{\rm coh,2}=l_{\rm coh}/(100~\rm kpc)$ \citep[e.g.,][]{ryu+08}. 
We consider that the CRs are in cosmic filaments and/or the galaxy groups with Kolmogorov turbulence, and ignore the cosmic expansion.
In addition, our Galaxy is located in the local group,
where the magnetic fields are probably stronger than the usual IGM.
These magnetic fields can potentially reduce the UHECR flux of $E_p\sim 10^{19}$ eV arriving at the Earth \citep{tmd14}. 
We should take the effects of these magnetic fields into account to discuss the arrival CR flux in detail. 
   
The escaping protons would diffuse in host galaxies of LLAGN,
and interact with gas in the interstellar medium (ISM) inside the galaxies.
The pion production efficiency of $pp$ inelastic collisions in the ISM is estimated to be $f_{\pi,\rm gal}\simeq K_{pp}n_{p,\rm gal}\sigma_{pp} c t_{\rm trap}\sim 8\times{10}^{-4} E_{p,16}^{-0.3}$, where $n_{p,\rm gal} \sim 1 ~\rm cm^{-3}$ is the mean nucleon density in the host galaxy, $t_{\rm trap} = h^2/4\kappa$ is the trapping time in the galaxy. 
We use the scale height $h\sim 1$ kpc and the diffusion coefficient estimated in our Galaxy, $\kappa\sim 3\times 10^{28}(E_p/1{\rm GeV})^{0.3} \rm~ cm^2~ s^{-1}$.
Note that the central regions of galaxies have much denser gases than the mean values of ISM that we use for the estimation.
The interaction of escaping protons in such dense cores of galaxies might be important.
The escaping protons are expected to be confined in IGM.
These protons are likely to interact with the protons or photons.
The efficiency of pion production in IGM is not low,
typically $\sim 10^{-2}$ below 100 PeV \citep{mal13},
which is likely to be more important than the reactions in ISM.
These processes might affect the diffuse neutrino flux.

\subsection{Constraints on neutron loading in the jet}

\citet{tt12} proposed a mass loading model to relativistic jets by relativistic neutrons made in the accretion flows. 
They consider that the relativistic neutrons whose Lorentz factor $\gamma_n \sim 3$ decaying at the polar region of a SMBH are able to provide the jets with some amount of mass and energy. 
They estimated that the relativistic neutrons can inject the energy of $\ljet \lesssim 2\times 10^{-3} \dot M c^2$
and the mass of $\mdjet \lesssim 4\times 10^{-4}\dot M$. 
This estimate results from the assumption of $L_n \sim 0.03 \dot M c^2$,
where $L_n$ is the total luminosity of neutrons from the accretion flow. 
The total luminosity of neutrons in our model is estimated as
\begin{equation}
 L_n \sim f_n \etacr \dot M c^2,
\end{equation}
where $f_n$ is the neutron generation efficiency.
The neutron generation efficiency is the same order of the pion production efficiency,
$f_n \sim f_\pi \lesssim 0.2$. 
From the fitting of the diffuse neutrino flux, we obtain $\etacr \sim 0.01$.
These results restrict $L_n \lesssim 2\times 10^{-3} \dot M c^2 $,
which is much lower than their assumption. 
In addition, resultant spectra of relativistic protons that are accelerated via stochastic acceleration are quite hard. 
This causes the differential luminosity and mass of the neutrons with $\gamma_n\sim 3$
to be much lower than the above restriction. 
Therefore, the neutron mass loading model is disfavored
when high-energy neutrinos are produced and limited by the observed neutrino data.

If LLAGN cannot accelerate the CR protons up to sufficiently high energy, 
the neutron injection model is not restricted from the neutrino observation.
For example, for the models with $\zeta \le 0.03$ and $q=5/3$, LLAGN cannot emit the neutrinos of $E_\nu \gtrsim 30$ TeV.
However, it is still not easy to achieve the required value of $f_n \etacr \sim 0.03$.
One reason is that $\etacr$ should be less than 0.1 in order to keep the structure of the RIAFs \citep{ktt14}.
Another reason is that the nature of collisionless plasma requires that the density should be low and limit $f_n \lesssim 0.3$.
Thus, we need an optimized situation for neutron generation
in order that the neutron injection model works as a jet mass loading mechanism.

\section{Discussion}\label{sec:discussion}

\subsection{Comparison to other AGN models}

In this work, we considered one of the AGN core models, in which CR acceleration and neutrino production occur in the vicinity of SMBH.  
Contrary to this work, in the previous literature, CR acceleration around the standard thin disk is assumed. However, since the disk plasma is typically collisional, faster dissipation is needed.  The shock in accretion flows \citep[e.g.,][]{pk83,ste+91} or between blobs \citep{am04}, and electric field acceleration \citep{kst14} have been speculated as underlying acceleration mechanisms.  
The acceleration mechanism at such inner regions is very uncertain.  For the efficient shock acceleration mechanism to work, $\tau_T\lesssim1$ is required to have collisionless shocks unmediated by radiation \citep[e.g.,][]{mi13}, but the condition depends on the radius and accretion rate. It is highly uncertain if electric field acceleration occurs since the gap formation may be prohibited by a copious plasma supplied from the disk to SMBH.  In any case, if one allows acceleration of CRs in the vicinity of disk, they should interact with ultraviolet photons supplied by multi-color blackbody emission from the standard disk \citep{ss73}.
Then, using the disk temperature around the innermost stable orbit $T_{\rm max}$, the typical neutrino energy is estimated to be
\begin{eqnarray}
E_\nu\approx0.05\frac{0.5m_pc^2\epeak}{\kb T_{\rm max}}\sim400~{\rm TeV}~\left(\frac{\kb T_{\rm max}}{20~{\rm eV}}\right)^{-1}.
\end{eqnarray}
Hence, for an averaged accretion disk spectrum observed in quasars, a suppression around sub-PeV energies is expected \citep{dmi14}.  In principle, it is possible to have lower-energy neutrinos by assuming high-temperature disks ad hoc.  However, in such models, all the relevant parameters (the CR normalization, spectral index, maximum energy and disk temperature) are essentially free parameters. Also, since gamma rays should not escape because of the large optical depths for pair production, these models should be regraded as {\it hidden neutrino source models}.  Note that the neutrino luminosity will be higher for AGN with higher disk luminosities.  Then, the well-observed X-ray luminosity function \citep{ued+03} suggests that neutrino emission is dominated by AGN with $L_X\gtrsim{10}^{44}~{\rm erg}~{\rm s}^{-1}$ \citep{mid14}.   

In the vicinity of the standard disk, $p\gamma$ interactions are usually the most important process \citep[e.g.,][]{sp94,am04}.  Also, the heavy jet has a problem in its energetics \citep{ad03,mid14}.  On the other hand, there are some discussions on $pp$ scenarios in radio galaxies \citep{bec+14}, assuming the existence of dense knots with $N_{H}\sim{10}^{24}~{\rm cm}^{-2}$.  If CRs are supplied by jets, the efficient neutrino production would significantly be diluted by their volume filling factor.  Also, note that steep spectra $s_\nu\gtrsim2.2$ are already ruled out by the multi-messenger data \citep{mal13}.  In principle, this can be avoided by requiring that GeV--TeV gamma rays are attenuated, where this model should be regarded as one of the hidden neutrino source models that are difficult to test.   

The most popular possibility is neutrino production in inner jets \citep[e.g.,][]{man95,ad01,mp01}.  If one adopts the simple one-zone model, most important contributions come from quasar-hosted blazars, and external radiation fields are the most important \citep{mid14,dmi14}. Whereas it is possible to explain ultra-high-energy CRs with heavy nuclei, a power-law CR spectrum is inconsistent with the absence of $\gg2$~PeV neutrinos.  To explain the IceCube data around PeV energies, the maximum energy of CRs has to be lower than ultra-high energies and another component is needed at low energies. On the other hand, \citet{tgg14} showed that, if a two-component model is invoked, BL Lac objects can be efficient emitters of PeV neutrinos without contradicting the observations of CRs. This is because the relative velocity between the spine and sheath allows us to have suitable target photon energies.  However, it is not clear how BL Lacs can make a dominant contribution to the diffuse neutrino efficiency, compared to that from quasar-hosted blazars.   

Among large scale jets, jets of Fanaroff-Riley II galaxies produce a non-relativistic cocoon shock and hot spot, and the latter is often bright at radio bands. The intergalactic density is usually too low to expect many neutrinos. However, since most AGN are expected to be located in galaxy clusters and groups, their contributions may be relevant \citep[e.g.,][]{min08,kot+09}. This possibility can be regarded as one of the $pp$ scenarios, which has been constrained by multi-messenger data \citep{mal13}.

\subsection{Gamma rays from RIAFs}\label{sec:gamma}

If neutrinos are produced by pion decay, gamma rays are also inevitably produced.
The generated spectrum and luminosity of these gamma rays are similar to those of the neutrinos.
However, high-energy gamma rays are absorbed by soft photons through $\gamma+\gamma\rightarrow e^+ + e^-$, so that the observed spectra of the gamma rays can be different from those of the neutrinos.   
In our model, internal absorption inside sources is relevant, and electromagnetic cascades are initiated.  
The emergent spectra are expected to have a break at the energy where the optical depth of pair production
$\tau_{\gamma\gamma} =1$.
We estimate the optical depth by
\begin{eqnarray}
\tau_{\gamma\gamma}(E_\gamma) \sim 0.2 \sigmat R N_\gamma \left(\epsilon_t\right) \epsilon_t, 
\end{eqnarray}
where $\epsilon_t \simeq (m_e c^2)^2/E_\gamma$ is the energy of the soft photons \citep[e.g.,][]{cb90}.
In our model, bright LLAGN, such as A3 model,
have the cutoff energy $\ecut\sim 5.4 $ GeV, while faint ones, such as A2 model, have $\ecut\sim 24$ GeV, as tabulated in Table \ref{tab:single}.
This means that bright LLAGN emit only multi-GeV photons and cannot emit TeV photons.
In this subsection, although we defer detailed studies of cascade emission, we here give the order-of-magnitude estimate on 
expected gamma-ray signatures, which may serve as tests of the LLAGN model.

Recent observations by {\it Fermi} show that the diffuse isotropic $\gamma$-ray background (IGB) intensity is $\sim 5\times10^{-7}~\rm GeV~cm^{-2}~s^{-1}~sr^{-1}$ at $E_\gamma \sim 1$ GeV, and $\sim 7\times10^{-8}~\rm GeV~cm^{-2}~s^{-1}~sr^{-1}$ at $E_\gamma \sim 100$ GeV \citep{fermi14igb}.  
This sets model-independent strong bounds on $pp$ scenarios \citep{mal13}, and spectral indices should be harder than $s_\nu\sim2.2$. 
The LLAGN model can avoid these constraints due to two reasons. First, the neutrino spectrum is harder than $s_\nu\sim2.0$, since stochastic acceleration or magnetic reconnection mechanisms predict hard spectra compared to the diffusive shock acceleration mechanism. Thus, direct gamma rays do not contribute to the IGB. Second, GeV--TeV gamma rays may not escape, and LLAGN can be regarded as hidden neutrino sources.  In reality, the situation depends on $\dot m$, and GeV--TeV gamma rays can be produced via cascades.  The maximum GeV--TeV flux can be estimated by assuming that all gamma rays escape and get cascaded in intergalactic space. Even in this case, noting that the gamma-ray flux is comparable to the neutrino flux, the estimated IGB flux is expected to be $\lesssim 10^{-7} ~\rm GeV~cm^{-2}~s^{-1}~sr^{-1}$ for B2 and B4, and $\lesssim 3\times 10^{-8} ~\rm GeV~cm^{-2}~s^{-1}~sr^{-1}$ for B1 and B3. However, in RIAFs, pairs produced via $\gamma+\gamma\rightarrow e^+ + e^-$ would lose mainly via synchrotron emission rather than inverse-Compton emission, so the IGB at sub-TeV energies is expected to be sufficiently lower than the observed intensity. More accurate calculation and examine the contribution to the IGB remains as a future work. 

How could we test the LLAGN model presented here?  Unfortunately, the averaged neutrino luminosity per source is too dim to detect individual sources.  Gamma-ray detections are also difficult although we expect that faint LLAGN can emit TeV photons. One of the examples of LLAGN with RIAFs is Sgr A* at the Galactic Center,
where thermal electrons at the inner region emit the radio band photons of $L_\gamma\sim 10^{36} ~\rm erg~s^{-1}$ at $E_\gamma \sim 4\times 10^{-3}$ eV \citep{yqn03}.
We calculate the inner part of the accretion flow of Sgr A* such that our model accounts for the observed radio luminosity. 
We use the same parameters as those for model A1 except for $\mbh=4\times10^6\msun$ and $\dot m = 1\times10^{-6}$.
Then, we find that protons are accelerated up to 10 TeV. 
The differential proton luminosity in this model is $E_p L_{E_p} \sim 1\times 10^{36}\rm ~erg~s^{-1}$ at $E_p \sim$ 10 TeV and $E_p L_{E_p} \sim 5\times 10^{32}\rm ~erg~s^{-1}$ at $E_p \sim$ 10 GeV.
These escaping protons are expected to emit GeV--TeV gamma rays via $pp$ reaction with surrounding materials, but this luminosity seems too low to explain the observed GeV--TeV gamma-ray fluxes \citep{liu+06,che+11}. 
These protons do not contribute to the observed flux of CR protons either.
The neutrino flux from Sgr A* in this model is $E_\nu F_{E_\nu} \sim 7\times 10^{-17}\rm ~erg~cm^{-2}~s^{-1}$ with peak energy $E_\nu\sim 0.3$ TeV, which is too faint to be observed. This accretion flow is so faint that gamma rays from neutral pion decay can escape from the flow directly. The gamma-ray flux is the same order of magnitude with that of neutrinos with peak energy $E_\gamma\sim $0.7 TeV, which is much lower than the gamma-ray flux at the Galactic Center observed by High Energy Stereoscopic System (HESS) \citep{hess09gc}.
Thus, this model does not contradict the observation of the Galactic Center and CR experiments.

Possibly, relatively brighter LLAGN might be able to be observed at GeV gamma rays. The cores of Cen A and M87 are candidates, and both GeV and TeV gamma rays are detected \citep{fermi09sl,hess06m87,hess09cenA,sah+13}.
For Cen A, the GeV gamma-ray spectrum cannot be smoothly connected to the TeV spectrum, and a break around 3 GeV has been suggested.
This could indicate the existence of two components.  Whereas some contributions could come from RIAFs, emission from jets are prominent \citep{tak11}, and it would not be easy to identify the RIAF component in observed spectra. It would be better to look for radio-quiet AGN with RIAFs that do not have strong jets.
We check the detectability of gamma rays from a nearby LLAGN.
NGC 3031 (M81) locates at $d\simeq 3.6$ Mpc and  has $L_{\rm bol}\sim 2\times 10^{41} ~ \rm erg~ s^{-1}$ with $\mbh \sim6\times 10^7 M_{\odot}$ \citep{ehf10}.
This luminosity can be obtained by our model with reference parameters except $\mbh=6.3\times 10^{7}M_{\odot}$.
The peak neutrino flux in this model is $E_\nu F_{E_\nu}\sim 7\times 10^{-10} \rm~GeV~cm^{-2}~s^{-1}$ at $E_\nu\sim 0.15 PeV$, which is too dim to be detected by IceCube.
The pair-production cutoff energy for this model is $\ecut \sim 6$ GeV, so that multi-GeV gamma rays are expected to escape.
The gamma-ray flux from M81 is at most the same order with that of neutrinos, $E_\gamma F_{E_\gamma}\sim 1\times 10^{-12} \rm~erg~cm^{-2}~s^{-1}$.
Thus, M81 could be detectable by {\it Fermi} or Cherenkov telescopes with low thresholds such as Major Atmospheric Gamma-ray Imaging Cherenkov Telescope (MAGIC) and Cherenkov Telescope Array (CTA), although the estimate above is too simple to predict the correct flux. 
Multimessenger studies are relevant to test the model and more precise calculations of photon spectra will be presented as a future work.

\section{Summary}\label{sec:summary}

We studied particle acceleration and associated neutrino emission from RIAFs of LLAGN.  
Various acceleration mechanisms have been suggested.  In this work, for demonstration, we consider stochastic acceleration, for which we can calculate spectra of escaping particles by solving the Fokker-Planck equation.  We modeled target photon fields in RIAFs by calculating inverse-Compton emission, based on the one-zone approximation.  Then we compared acceleration, escape, and cooling time scales,
and found that in LLAGN, proton acceleration is typically suppressed by diffusive escape rather than cooling processes. We also found that LLAGN can have the protons up to more than 10 PeV for reasonable ranges of $\dot m$, $\mbh$, $\zeta$ and $q$.
Then, the $pp$ or $p\gamma$ production may lead to PeV neutrinos.  Note that production of ultra-high-energy CRs is not expected in this model. 
The CR acceleration efficiency is highly uncertain, so we treated it as a free parameter assuming that the total luminosity of escaping CRs is equal to $\eta_{\rm cr}{\dot M} c^2$, including both diffusive and advective escape.  Then, the luminosity of CRs escape via diffusion is estimated to be around $1\times10^{41}$ erg s$^{-1}$ in our reference model.  We calculate associated neutrino emission, and found that high-energy neutrino production occurs mainly via $pp$ interactions, and the meson production efficiency is typically the order of 1\%.  The neutrino spectrum is hard since CRs are assumed to be accelerated via the stochastic acceleration mechanism.  

We also calculated the diffuse neutrino intensity by using the $H_\alpha$ luminosity function of LLAGN and assuming no redshift evolution.  Interestingly, we found that the observed IceCube data can be fitted for reasonable parameters if $\sim1$\% of the accretion luminosity is carried by CRs. This fraction gurantees our assumption that the CRs do not affect the dynamical structure of RIAFs \citep{ktt14}. The number density of LLAGN is the order of $\sim{10}^{-3}-{10}^{-2}~{\rm Mpc}^{-3}$, which is much higher than those of radio-loud AGN including blazars.  Since the spectrum is hard, this result does not contradict the diffuse gamma-ray bound \citep{mal13} and observed CR flux. 

Whereas RIAFs of LLAGN can provide interesting targets of high-energy neutrino and gamma-ray observations, unfortunately, there are many uncertainties in the model.  First, parameters related to acceleration are uncertain, although values we adopt are often used in the different literature such as gamma-ray bursts. However, although we considered stochastic acceleration, one should keep in mind that our neutrino flux calculations can be applied to different possibilities such as acceleration via magnetic reconnections. For more reliable predictions, we need better knowledge on the distribution of non-thermal particles, which could be achieved by future particle-in-cell simulations.  Second, the luminosity function of LLAGN is quite uncertain due to their faintness.  Obviously, to estimate the diffuse neutrino intensity, more observational data on the shape of the luminosity function in the faint end and their redshift evolution are needed, as well as the mass function of SMBHs hosted by LLAGN.  In addition, contributions from RIAFs with the critical mass accretion rate, at which RIAFs change to the standard disk, may also be relevant.  This implies the importance of understanding the physical relationship between LLAGN and Seyferts. 

One of the potentially interesting points of the LLAGN model is that one could explain the latest IceCube data around 10 TeV.  The latest data suggest steeper indices of $s_\nu\sim2.3-2.5$, which seems challenging for many models.  
Galactic sources may be responsible for $\lesssim100$~TeV neutrinos, but it is premature to discuss such a two-component scenario due to the lack of compelling anisotropy.  It could be explained by an exponential cutoff or spectral break of starburst galaxies, but hard indices of $s_\nu\sim2$ are needed
\citep{sen+15}.
Alternatively, hidden neutrino sources can provide viable possibilities. Such a speculation includes not only the AGN core models including the LLAGN model but also orphan neutrino production in low-power gamma-ray burst jets \citep{mi13}.     

As a final remark, we stress that the neutrino observations may be powerful for proving physics of accretion disks and jets. In this work, we calculated the neutron generation rate in RIAF \citep[see also][]{ktt14}, and argue that an optimized parameter values are required for RIAFs to have as high neutron generation rate as suggested by \citet{tt12}. As long as CR spectral indices are hard as expected in stochastic acceleration or magnetic reconnection, the model of jet mass loading mediated by neutrons is strongly restricted by neutrino observations unless $\zeta$ is very low. 


\acknowledgments

We thank Katsuaki Asano, Jun Kakuwa, Matt Kistler, Fumio Takahara, and Shuta Tanaka for fruitful discussion. 
S.S.K. thanks Yutaka Fujita, Kentaro Nagamine, and Hideyuki Tagoshi for continuous encouragement. 
K. M. thanks Rashid Sunyaev for discussion about the disk photon field. 
This work is partly supported by Grant-in-Aid for JSPS Fellowships No. 251784 (S.S.K.). 
This work is supported by NASA through Hubble Fellowship Grant No. 51310.01 awarded by the STScI, 
which is operated by the Association of Universities for Research in Astronomy, Inc., for NASA, under Contract No. NAS 5-26555 (K.M.) 
The result of this work was presented by K. M. at the JSI workshop in November 2014.   

\appendix

\section{Calculation method for the spectrum from thermal electrons}\label{app:photon}

\subsection{Synchrotron and Bremsstrahlung}

We use a fitting formula for the synchrotron and bremsstrahlung \citep{ny95}.
Hear, we use the cgs units. 
The cooling rates by bremsstrahlung of electron-electron $q_{ei}$ and ion-electron $q_{ee}$ are represented as follows,
\begin{equation}
 q_{ei} =1.48 \times 10^{-22} n_e^2 F_{ei}(\theta_e), 
\end{equation}
\begin{eqnarray}
 q_{ee}=  \left\{
   \begin{array}{ll}
    2.56\times 10^{-22} n_e^2\theta_e^{3/2}\left(1+1.1\theta_e + \theta_e^2 -1.25\theta_e^{5/2}\right) & (\theta_e < 1) \\
    3.40\times 10^{-22} n_e^2\theta_e\left[\ln(1.123\theta_e + 1.28)\right] & (\theta_e > 1)\\
   \end{array} \right. ,
\end{eqnarray} 
where we use 
\begin{eqnarray} 
 F_{ei}=  \left\{
   \begin{array}{ll}
    4\left(\frac{2\theta_e}{\pi^3}\right)^{0.5}(1+1.781\theta_e^{1.34}) & (\theta_e < 1) \\
    \frac{9\theta_e}{2\pi}[\ln(1.123\theta_e + 0.48)+1.5] & (\theta_e > 1) \\
   \end{array} \right. .
\end{eqnarray} 
The emissivity of bremsstrahlung is related to the cooling rates as
\begin{equation}
 q_{ei} + q_{ee} = \int_0^{\infty}d\nu j_{\nu,\rm br}. 
\end{equation}
Approximating $j_{\nu,\rm br}= j_0 \exp(-h\nu/(\kb T_e))$, where $j_0$ does not depend on $\nu$, 
we can write 
\begin{equation}
 j_{\nu,\rm br} = q_{\rm br} \frac{h}{\kb T_e} \exp\left(-\frac{h\nu}{\kb T_e}\right)
\end{equation}
where $q_{\rm br} = q_{ei}+q_{ee}$ is the cooling rate per unit volume.  
We assume the gaunt factor is unity.

The synchrotron emissivity is 
\begin{equation}
 j_{\nu,sy} = \frac{4\pi e^2n_e \nu}{\sqrt{3} c K_2(1/\theta_e)} I'\left(\frac{4\pi m_e c\nu}{3 e B \theta_e^2}\right) ,
\end{equation}
where
\begin{equation}
 I'(x) = \frac{4.0505}{x^{1/6}}\left(1+\frac{0.4}{x^{1/4}} + \frac{0.5316}{x^{1/2}}\right)\exp(-1.8899x^{1/3}). 
\end{equation}

\subsection{Radiative transfer}

We define the optical depth for absorption as
\begin{equation}
 \tau_\nu\equiv \frac{\sqrt \pi}{2} \kappa_\nu H \sim \frac{\sqrt \pi}{2} \kappa_\nu R,
\end{equation}
where
\begin{equation}
 \kappa_\nu = \frac{j_{\nu,br} + j_{\nu,sy}}{4\pi B_\nu}
\end{equation}
is the absorption coefficient and $B_\nu$ is the Plank function. 
The photon energy flux from a RIAF is represented as \citep{mmk97}
\begin{equation}
 F_\nu = \frac{2\pi}{\sqrt 3}B_\nu \left[1-\exp(-2\sqrt 3 \tau_\nu)\right], 
\end{equation}
where we use Eddington approximation \citep{rl79} when estimating the vertical energy flux.
The luminosity by the synchrotron and bremsstrahlung is estimated as
\begin{equation}
 L_{\nu,0} = 2\pi R^2 F_\nu.
\end{equation}

\subsection{Inverse Compton scattering}

We calculate the spectrum of the inverse Compton scattering.
Seed photons are the photon field by bremsstrahlung and synchrotron.
Assuming homogeneous and isotropic distribution, the photon occupation number evolves by  \citep[cf.][]{cb90},
\begin{eqnarray}
 \frac{dN_{\gamma}(\epsilon)}{dt}\Delta\epsilon = 
  - N_{\gamma}(\epsilon)\Delta\epsilon\int d\gamma N_e(\gamma)R_c(\epsilon,\gamma)  \nonumber \\
  + \int d\gamma \int d\epsilon' N_e(\gamma)N_{\gamma}(\epsilon')R_c(\epsilon',\gamma) P_c(\epsilon;\epsilon',\gamma)
  + \dot N_{\gamma,0}\Delta\epsilon - \dot N_{\gamma,esc}\Delta\epsilon, \label{eq:photon}
\end{eqnarray}
where $\epsilon = h\nu/(m_e c^2)$, $N_{\gamma}(\epsilon)$ is the differential number density of photon,
$R_c(\epsilon,\gamma) \rm [ cm^3 ~s^{-1}]$ is the reaction rate of the electrons of Lorentz factor $\gamma$ 
and the photons of energy $\epsilon$, 
and $P_c(\epsilon;\epsilon',\gamma)$ is the probability that the reactions by the photons of energy $\epsilon'$ 
and electrons of energy $\gamma$ create the photons of energy $\epsilon$.
We add the injection rate of seed photons $\dot N_{\gamma,0}$ and escape rate of photons $\dot N_{\gamma,esc} = N_\gamma(\epsilon)/(R/c)$.
We calculate the steady state solution of this equation.
The differential number density can be expanded by the number of Compton scattering as 
$N_\gamma(\epsilon)=N_{\gamma,0} + N_{\gamma,1} + ...$, 
where $N_{\gamma,0}$ is determine by the balance of the escape and the injection 
\begin{equation}
 N_{\gamma,0}/(R/c) =(1-\tau_{\rm T}) \dot N_{\gamma,0} = L_{\epsilon,0}/(\pi R^3 h \nu),
\end{equation}
where $L_{\epsilon,0} = (m_e c^2/h) L_{\nu,0}$.
Since optical depth of the flow is less than unity, 
we approximate the effect of first term of right-hand side of Equation (\ref{eq:photon})
by multiplying the factor $1-\tau_{\rm T}$.
We obtain the $N_{\gamma,i}$ by solving the following equation,
\begin{equation}
 \frac{N_{\gamma,i}(\epsilon)}{R/c} \Delta\epsilon
  = (1-\tau_{\rm T})\int d\gamma \int d\epsilon'  N_e(\gamma)N_{\gamma,i-1}(\epsilon') R_c(\epsilon',\gamma) P_c(\epsilon;\epsilon',\gamma), 
\end{equation}
where we use the approximation of $(1-\tau_{\rm T})$ again.
We use the fitting formula for $R_c$ \citep{cb90}
\begin{equation}
 R(\epsilon,\gamma) = \frac{3 c \sigma_T}{8\gamma\epsilon}\left[\left(1-\frac{2}{\gamma\epsilon}-\frac{2}{\gamma^2\epsilon^2}
\right)\ln\left(1+2\gamma\epsilon\right) + \frac{1}{2}+\frac{4}{\gamma\epsilon} - \frac{1}{2(1+2\gamma\epsilon)^2}\right]. 
\end{equation}
From the energy conservation, the reaction with $4\gamma^2\epsilon/3 > \gamma + \epsilon $ does not occur. 
In this case, we take $R(\epsilon,\gamma)=0$. 
We use delta function approximation for $P_c$, 
\begin{equation}
 N_\gamma(\epsilon')P_c(\epsilon;\epsilon',\gamma) = N_\gamma(\epsilon')\Delta\epsilon'\delta\left(\epsilon - \frac{4}{3}\gamma^2\epsilon' \right). 
\end{equation}
From the treatment described above, we obtain 
\begin{equation}
 N_{\gamma,i}(\epsilon) = 
\frac R c \int d\gamma \frac{3}{4\gamma^2}  N_e(\gamma)N_{\gamma,i-1}\left(\frac{3\epsilon}{4\gamma^2}\right) 
R_c\left(\frac{3\epsilon}{4\gamma^2},\gamma\right). 
\end{equation}
We calculate $N_{\gamma,i}(\epsilon)$
until
the number density of the $i$ times scattered photons is much less than that of the bremsstrahlung in all the range in which we are interested.


\clearpage


\end{document}